# Material Independent Long Distance Pulling, Trapping, and Rotation of Fully Immersed Multiple Objects with a Single Optical Set-up

Md. Masudur Rahman[1], Ayed Al Sayem[1], M. R. C. Mahdy[2*], Md. Ehsanul Haque[1], Rakibul Islam[1], S. Tanvir-ur-Rahman Chowdhury [1], Manuel Nieto-Vesperinas[3*] and Md. Abdul Matin[1]

[1]*Department of EEE, Bangladesh University of Engineering & Technology, Dhaka, Bangladesh*

[2]*Department of Electrical and Computer Engineering (ECE), National University of Singapore, Singapore*

[3] *Instituto de Ciencia de Materiales de Madrid, Consejo Superior de Investigaciones Cientificas, Campus de Cantoblanco, 28049, Madrid, Spain*

∗ *Corresponding Authors' E-mails: A0107276@nus.edu.sg*

*and* mnieto@icmm.csic.es

**Abstract: Optical pulling with tractor beams is so far highly dependent on (i) the property of embedding background or the particle itslef , (ii) the number of the particles and/or (iii) the manual ramping of beam phase. A necessary theoretical solution of these problems is proposed here. This article demonstrates a novel active tractor beam for multiple fully immersed objects with its additional abilities of yielding a controlled rotation and a desired 3D trapping. Continuous and stable long distance levitation, controlled rotation and 3D trapping are demonstrated with a single optical set-up by using two coaxial, or even non-coaxial, superimposed non-diffracting *higher order Bessel beams of reverse helical nature and different frequencies*. The superimposed beam has periodic intensity variations both along and around the beam-axis because of the difference in longitudinal wave-vectors and beam orders, respectively. The difference in frequencies of two laser beams makes the intensity pattern move along and around the beam-axis in a continuous way without manual ramping of phase, which allows for either linear motion (forward or backward) or angular movement (clockwise or anticlockwise) of fully immersed multiple particles. As a major contribution, the condition for increasing the target binding regions is also proposed to manipulate multiple immersed objects of different sizes and shapes.**

## INTRODUCTION

Optical manipulation and optical forces play a substantial role in various areas of physics, bio-physics [1- 4] and in medical applications [5-9]. Optical forces have been used in a variety of ways such as optical cooling in atomic physics [10], demonstrating chaotic quivering of micro-cavities [11], optical cooling of mechanical modes [12-14], MEMs and nano-opto-mechanical systems [15, 16], to name a few. Recently, optical forces have also been introduced in two dimensional materials such as graphene [17]. Trapping and manipulation with optical tweezers was started by Ashkin long ago [18] and it has served as a rudimentary tool in medical and biological applications [5- 9]. Tweezers have been used in a variety of applications including the development of basic understanding of the properties of biological polymers [19, 20] and the development of low-drift trapping instrumentation [1-3].

Many experiments in the areas of optical manipulation [21-25] involved objects immersed in a non-vacuum environment: e.g., gels, solutions, gases, water or dielectrics in general. However, both a three-dimensional trapping [26, 27] and controlled rotation [28-31] of fully immersed multiple particles, (e.g., immersed in gels, solutions, highly dense gases, water or dielectrics), at desired positions with a single setup of light source, are still missing in the literature from both a theoretical and experimental point of view. With the emerging techniques of tractor beams [21, 24, 32], a compact theoretical solution for all kinds of optical manipulations by using the current existing technology, should constitute a breakthrough towards pushing manipulation much forward.

Previous works regarding optical pulling using tractor beams can be divided into well-defined procedures which mainly are based on: (a) the use of a single non-paraxial Bessel beam [33, 34], (b) building a standing wave named the optical conveyor-belt [35, 36], (c) on employing gain background media [37], (d) generating a negative force by exciting multipoles inside a scatterer [38], (e) placing the particle between two different media to make linearly polarized light [39] to act as a tractor beam and (f) introducing the optical conveyor [40] as a travelling wave.

In most works so far, optical pulling is based on the scattering force, and long distance pulling is not possible [33, 34, 38, 39]. In Ref. [41] a solenoidal beam has been introduced to locate a particle by using the gradient force, and to spirally shift the particle with the help of radiation pressure. Only in a very recent paper, a pulling phenomenon at a moderately good speed, (higher than above quoted cases), has been shown by using the photophoretic force [42] for artificially coated objects. Neither dielectric Rayleigh particles, nor those having a radius slightly larger than that of Mie ones, can be pulled based on the techniques reported in [33-34, 38]. A conveyor operation by two co-propagating zero order Bessel beams has been described in [40] only for a single particle embedded in a single medium using a manual ramping of the phase, which is not suitable for multiple particle manipulation. All aforementioned works possess some core inherent complications and real world applications (such as biological, or in life or in space research [24, 25, 43, 44]) of tractor beams are still far away to reach.

In order to try to step in the next revolution in optical manipulation, a theoretical foundation for almost all kinds of feasible optical manipulation procedures (e.g. long distance pulling, controlled rotation and trapping at any desired position of fully immersed multiple particles) inside a material background medium, (either homogenous or inhomogeneous), with a single optical set-up, is the ultimate goal of this work.We propose here a multiple particle manipulation mechanism, different from [41], and capable of yielding almost all kinds of manipulation configurations, including trapping, rotation and levitation.

Henceforth, by using superposition of two higher order co-axial Bessel beams, (with opposite order of each other, with two different frequencies and with different cone angles), we theoretically demonstrate almost all feasible optical manipulations for fully immersed particles in a continuous and controlled manner and without using manual ramping of phase [40]; [cf. Figs. 1(a) and 1(b)]. To this end, we have considered some critical characteristics of a Bessel beam-pair such as a choice of higher but reverse orders and different frequencies. Moreover, we have shown that those two incident Bessel beams are not necessarily required to be co-axial (cf.Fig. 1(c)).

In this work, by increasing the number of binding regions inside the liquid, we have demonstrated multiple particle manipulations such as: (i) conveying them in a longitudinal direction in a 'fully stable' path, (either pushing or pulling them with a constant velocity at will); (ii) submitting them to a stable rotation in an azimuthal direction, (either clockwise and anticlockwise); (iii) establishing necessary conditions of the angular and linear velocities for a stable conveying, and also for a controlled trapping at any desired position in the liquid medium; (iv) calculating those possible disturbing forces, (i.e., those viscous, gravitational and centrifugal), versus the optical gradient force. Our work paves the way to utilizing the highest number of feasible manipulation with a single setup, especially for biological and other real life applications [25, 43-45].

A. **A SINGLE BESSEL BEAM AND SUPERIMPOSED BEAMS: AIR TO LIQUID**

The full expression of the electric and magnetic fields of the incident Bessel beam in the air is [46]:

$$\begin{bmatrix} e_{z.g} \\ h_{z.g} \\ e_{\rho.g} \\ h_{\rho.g} \\ e_{\phi.g} \\ h_{\phi.g} \end{bmatrix} = \begin{bmatrix} E_{z.g} \\ H_{z.g} \\ E_{\rho.g} \\ H_{\rho.g} \\ E_{\phi.g} \\ H_{\phi.g} \end{bmatrix} e^{j(\beta_g z_g + m\phi_g - \omega t)} \text{, where} \begin{bmatrix} E_{z.g} \\ H_{z.g} \\ E_{\rho.g} \\ H_{\rho.g} \\ E_{\phi.g} \\ H_{\phi.g} \end{bmatrix} = \begin{bmatrix} C_E J_m(q_g \rho_g) \\ C_H J_m(q_g \rho_g) \\ -\frac{m\omega\mu_g}{\rho_g q_g^2} C_H J_m(q_g \rho_g) + j \cdot \frac{\beta_g}{q_g} C_E J_m'(q_g \rho_g) \\ +\frac{m\omega\epsilon_g}{\rho_g q_g^2} C_E J_m(q_g \rho_g) + j \cdot \frac{\beta_g}{q_g} C_H J_m'(q_g \rho_g) \\ -\frac{m\beta_g}{\rho_g q_g^2} C_E J_m(q_g \rho_g) - j \cdot \frac{\omega\mu_g}{q_g} C_H J_m'(q_g \rho_g) \\ -\frac{m\beta_g}{\rho_g q_g^2} C_H J_m(q_g \rho_g) + j \cdot \frac{\omega\epsilon_g}{q_g} C_E J_m'(q_g \rho_g) \end{bmatrix} \quad (1)$$

Here, $e_{\rho.g}$, $e_{\phi.g}$, $e_{z.g}$ and $h_{\rho.g}$, $h_{\phi.g}$, $h_{z.g}$ are instantaneous electric and magnetic fields in the air, $E_{\rho.g}$, $E_{\phi.g}$, $E_{z.g}$ and $H_{\rho.g}$, $H_{\phi.g}$, $H_{z.g}$are electric and magnetic fields in a cylindrical co-ordinates, respectively. $\beta_g$ and $q_g$ are wave-vectors in the longitudinal and radial directions in the air, $m$ is the Bessel beam order, $\omega$ is the angular frequency, $t$ is time, $\epsilon_g$ and $\mu_g$ are the permittivity and permeability of the air, respectively,$C_E$ and $C_H$ are the electric and magnetic field coefficients of the Bessel beam. $\rho_g, \phi_g, z_g$ are the cylindrical coordinates of the position vector in the air. However, the property of the transmitted beam inside water is not fully same as that of the incident beam (cf. Fig. 1(d)). A non-parxial single gradientless Bessel beam may not be capable of pulling a dipole object by overcoming the viscous and other disturbing forces.

As a result, strong gradient changes of intensity in the azimuthal direction are achieved using superposition of Bessel-beams with different orders sharing the same beam-axis (cf. Fig. 1 (a) and (b)). To make the system appropriately working, it is mandatory to use reverse orders of each other so that particles are not vulnerable to angular acceleration. If strong intensity gradients can be created in all directions, (azimuthal, longitudinal and radial) by using two superimposed Bessel beams, it is possible to bi-directionally shift, rotate, or trap multiple particles, independently of size and material properties, with proper control and full stability, even if the particles are fully immersed in a liquid. The two Bessel beams, however, should not necessarily be co-axial. An angle difference between the two beams can serve as a new degree of freedom to levitate multiple fully immersed particles from a specific depth. The schematic figure of the non-coaxial setup is shown in Fig. 1(c).

B. **PHYSICAL MECHANISM OF GENERALIZED PARTICLE MANIPULATION**

Rather than depending on the scattering optical force to pull a particle, it is imperative to use a gradient force to serve the same purpose, because the latter is much larger than the former [34], and it will work for any particle having a dielectric constant $\epsilon_p$different from that of surrounding medium $\epsilon_m$. The components of the gradient force, (cf. Eqs.(22s)-(25s) in the supplement), in cylindrical coordinates are given by [47-52]:

$$\begin{bmatrix} {}_{grad}^{\;z}F \\ {}_{grad}^{\;\rho}F \\ {}_{grad}^{\;\phi}F \end{bmatrix} = \frac{\eta_m}{2\,c}\left(3\,\frac{\eta_p^2-\eta_m^2}{\eta_p^2+2\eta_m^2}\right)\begin{bmatrix} \frac{\partial}{\partial z}I \\ \frac{\partial}{\partial \rho}I \\ \frac{\partial}{\rho\partial \phi}I \end{bmatrix} V\,,\ \text{ where } V = \frac{4}{3}\pi r_p^3 \text{ and } I = \eta_m \epsilon_0 c\left(\frac{1}{\sqrt{2}}|E|\right)^2 \tag{2}$$

Here, $\eta_m$ and $\eta_p$ stand for the refractive index of the medium and the particle, respectively; $c$is the light velocity in vacuum, $V$ is the volume of the particle, $r_p$ is the particle-radius, ${}_{grad}^{\;\rho}F$, ${}_{grad}^{\;\phi}F$, ${}_{grad}^{\;z}F$ are the gradient forces in the $\rho, \phi, z$ directions, respectively, and $I$ is the total intensity of light at the point $(\rho, \phi, z)$.

Hence, the optical gradient force $\boldsymbol{F_{grad}}$ and potential $P$ are given as [50]:

$$\boldsymbol{F_{grad}} = {}_{grad}^{\;\rho}F\boldsymbol{\hat{\rho}} + {}_{grad}^{\;\phi}F\boldsymbol{\hat{\phi}} + {}_{grad}^{\;z}F\boldsymbol{\hat{z}} \tag{3}$$

$$P = \frac{\eta_m}{2\,c}\left(3\,\frac{\eta_p^2-\eta_m^2}{\eta_p^2+2\eta_m^2}\right)V\,(-I) \tag{4}$$

Fig.2a illustrates the binding and manipulation of a particle with the help of the gradient force. For a light-seeking particle, $\eta_p > \eta_m$. In this case, at any point at the left of intensity maxima the slope is positive and so is the gradient force, i.e. it is directed towards these maxima along the ($+z$ or $+\rho$ or $+\phi$)-axis. At any point at the right of the maxima, on the other hand, the slope is negative and the converse attraction towards those maxima along the ($-z$ or $-\rho$ or $-\phi$)-axis occur. As a result, a light-seeking particle will be fixed at a point near the nearest maxima where the total force is zero, (cf. Fig. 3s in the supplement), having the minimum potential, (cf. Fig. 2b). For a dark-seeking particle, however, $\eta_p < \eta_m$ and the particle will be fixed at a point near the nearest intensity minima. By controlling the intensity pattern in any direction, (longitudinal, azimuthal and radial), a particle can be manipulated in that direction in a controlled and stable fashion (Cf. Video File 1).

To ensure optical binding, the gradient force must be strong enough to counter-balance other disturbing forces acting on the particle [53]. If the gradient force peak is greater than the summation of those disturbing forces, the particle will be bound at a point where the net force is zero. That is the equilibrium position, (cf. Fig. 2a in this main article and Fig. 3s in the Supplement). Of course, no such equilibrium point exists if the resultant disturbing forces are larger than the gradient force peak.

### C. ALL POSSIBLE PARTICLE MANIPULATION OPERATIONS USING STRUCTURED BESSEL BEAMS

By superimposing two or more Bessel beams of slightly different characteristics, it is possible to create an intensity gradient in the three directions. Notice the difference of our work from the proposal of a time-dependent manual phase-shift with one of two Bessel-beams, (only of zero order) of Ref. [40]. In addition, in [40] no approach for multiple particle manipulations, 3D trapping and rotation is put forward. Also, on using a concept similar to that of [40] and just by using higher Bessel orders, (cf. Fig. 2s(b) in the supplement), one cannot achieve multiple particle manipulation without some critical characteristics, (specially the reverse orders and the certain difference of few specific characteristics) of the Bessel beam pair as considered here. We propose that by maintaining a certain difference of characteristics (i.e. difference in frequency and/or the paraxial angle) between the beams, multiple particles can be fully manipulated (i.e. by pulling, pushing, trapping, and rotation) in a continuous way, independently of their material and size. A detailed mathematical proof of our proposal is given in the Supplementary Section.

The components of the resultant electric field due to the superposition of the two beams of the Bessel pair are:

$$\begin{bmatrix} e_{z.l} \\ e_{\rho.l} \\ e_{\phi.l} \end{bmatrix} = \begin{bmatrix} E_{z.l}^{(1)} exp\left[j\left(\beta_l^{(1)} z_l + m^{(1)}\phi_l - \omega^{(1)} t\right)\right] + E_{z.l}^{(2)} exp\left[j\left(\beta_l^{(2)} z_l + m^{(2)}\phi_l - \omega^{(2)} t\right)\right] \\ E_{\rho.l}^{(1)} exp\left[j\left(\beta_l^{(1)} z_l + m^{(1)}\phi_l - \omega^{(1)} t\right)\right] + E_{\rho.l}^{(2)} exp\left[j\left(\beta_l^{(2)} z_l + m^{(2)}\phi_l - \omega^{(2)} t\right)\right] \\ E_{\phi.l}^{(1)} exp\left[j\left(\beta_l^{(1)} z_l + m^{(1)}\phi_l - \omega^{(1)} t\right)\right] + E_{\phi.l}^{(2)} exp\left[j\left(\beta_l^{(2)} z_l + m^{(2)}\phi_l - \omega^{(2)} t\right)\right] \end{bmatrix} \tag{5a}$$

Here $E_{\rho.l}^{(1)}, E_{\psi.l}^{(1)}, E_{z.l}^{(1)}$ are the electric field components of the first Bessel beam excluding the factor $exp\left[j\left(\beta_l^{(1)} z_l + m^{(1)}\phi_l - \omega^{(1)} t\right)\right]$ and $E_{\rho.l}^{(2)}, E_{\psi.l}^{(2)}, E_{z.l}^{(2)}$ represent the electric field components of the second Bessel beam excluding $exp\left[j\left(\beta_l^{(2)} z_l + m^{(2)}\phi_l - \omega^{(2)} t\right)\right]$ in cylindrical coordinates in the liquid medium, $\beta_l^{(1)}$ and $\beta_l^{(2)}$ are longitudinal wave-vectors in the liquid, $m^{(1)}$ and $m^{(2)}$ are the beam-orders, $\omega^{(1)}$ and $\omega^{(2)}$ stand for the angular frequencies of the first and second Bessel beams, respectively.

Considering the properties of the Bessel beam, and by rigorous calculations, the intensity of the resultant superimposed Bessel beams in the liquid medium, has been derived as, (cf. Supplementary Section 2.2 for details):

$$I_L = \eta\epsilon_0 c \begin{bmatrix} \frac{1}{2}\left|E_{\rho.l}^{(1)}\right|^2 + \frac{1}{2}\left|E_{\rho.l}^{(2)}\right|^2 + \left|E_{\rho.l}^{(1)}\right|\left|E_{\rho.l}^{(2)}\right| \cos\left(\left(\beta_l^{(1)} - \beta_l^{(2)}\right)z_l + \left(m^{(1)} - m^{(2)}\right)\phi_l - \left(\omega^{(1)} - \omega^{(2)}\right)t + \angle E_{\rho.l}^{(1)} - \angle E_{\rho.l}^{(2)}\right) \\ + \frac{1}{2}\left|E_{\psi.l}^{(1)}\right|^2 + \frac{1}{2}\left|E_{\psi.l}^{(2)}\right|^2 + \left|E_{\psi.l}^{(1)}\right|\left|E_{\psi.l}^{(2)}\right| \cos\left(\left(\beta_l^{(1)} - \beta_l^{(2)}\right)z_l + \left(m^{(1)} - m^{(2)}\right)\phi_l - \left(\omega^{(1)} - \omega^{(2)}\right)t + \angle E_{\psi.l}^{(1)} - \angle E_{\psi.l}^{(2)}\right) \\ + \frac{1}{2}\left|E_{z.l}^{(1)}\right|^2 + \frac{1}{2}\left|E_{z.l}^{(2)}\right|^2 + \left|E_{z.l}^{(1)}\right|\left|E_{z.l}^{(2)}\right| \cos\left(\left(\beta_l^{(1)} - \beta_l^{(2)}\right)z_l + \left(m^{(1)} - m^{(2)}\right)\phi_l - \left(\omega^{(1)} - \omega^{(2)}\right)t + \angle E_{z.l}^{(1)} - \angle E_{z.l}^{(2)}\right) \end{bmatrix} \quad (5b)$$

In Eq.(5b), the phases of the oscillating terms of the superimposed Bessel beam pair include" $\left(\beta_l^{(1)} - \beta_l^{(2)}\right)z_l + \left(m^{(1)} - m^{(2)}\right)\phi_l - \left(\omega^{(1)} - \omega^{(2)}\right)t$" which implies that the intensity gradient, and hence the change of the potential profile in the $z_l$-direction if $\beta_l^{(1)} \neq \beta_l^{(2)}$ (cf. Fig. 2d) and in the $\phi_l$-direction if $m^{(1)} \neq m^{(2)}$ (cf. Fig. 2c(iii)) can be simultaneously (cf. Fig. 2b(i)), or separately, achieved.The intensity pattern will not change with time if $\omega^{(1)} = \omega^{(2)}$, which means that the particle can be trapped at certain location using this condition. But the case $\omega^{(1)} \neq \omega^{(2)}$ along with $\beta_l^{(1)} \neq \beta_l^{(2)}$ and $m^{(1)} \neq m^{(2)}$ will offer a continuously moving intensity-pattern (and hence the continuous movement of the minimum potential regions) in the longitudinal direction at a linear speed of $v_{z.l}^{(L)} = \left(\omega^{(1)} - \omega^{(2)}\right)/\left(\beta_l^{(1)} - \beta_l^{(2)}\right)$, (cf. Eq.(33s) in the supplement), and in the azimuthal direction at a circular speed of $\Omega_\phi^{(C)} = \left(\omega^{(1)} - \omega^{(2)}\right)/\left(m^{(1)} - m^{(2)}\right)$ (cf. Eq.(37s) in the supplemnt and Fig. 3 in the main article). Thus the particle will also move to relocate itself at the newly located minimum potential region, (cf. Fig. 4s in the supplement). The azimuthal scattering components increase with the Bessel order [54]. On the other hand, the conveying of a particle in the azimuthal direction requires higher order Bessel beams of different orders. This problem involving the azimuthal scattering component, ultimately producing an angular acceleration, can easily be solved if a superimposed beam pair of reverse order is used. If $m^{(1)} = -m^{(2)}$and $\lambda^{(1)} \approx \lambda^{(2)}(= \lambda)$, $\left|\boldsymbol{E}^{(1)}\right| \approx \left|\boldsymbol{E}^{(2)}\right|$, then the azimuthal components of the scattering force of the superimposed beams are be found as follows.

$$\begin{bmatrix} (F_p)_\phi\big|_{(m^{(1)},m^{(2)})=(+l,-l)} \\ (F_m)_\phi\big|_{(m^{(1)},m^{(2)})=(+l,-l)} \\ (F_{pm})_\phi\big|_{(m^{(1)},m^{(2)})=(+l,-l)} \end{bmatrix}$$

$$\approx \begin{bmatrix} 0 \\ 0 \\ -\frac{1}{c} Re\left\{ a_{Mie}\, b_{Mie}^{*}\, \frac{1}{2} \begin{pmatrix} l\left(\cos 2\theta_{cone}^{(2)} - \cos 2\theta_{cone}^{(1)}\right)\frac{(\lambda/2\pi)}{\rho}(|C_E|^2 + |C_H|^2) J_l^2 \\ +j\left(\left(\sin 2\theta_{cone}^{(2)} + \sin 2\theta_{cone}^{(1)}\right) - \frac{1}{2}\left(\sin 4\theta_{cone}^{(2)} + \sin 4\theta_{cone}^{(1)}\right)\right) C_E C_H^{*} \frac{1}{2}(J_{l+1} - J_{l-1}) J_l \end{pmatrix} \right\} \end{bmatrix} \tag{5c}$$

Here, $a_{Mie}$ and $b_{Mie}$ are Mie-coefficients and $\theta_{cone}^{(1)}$ and $\theta_{cone}^{(2)}$ are the cone-angles of the first and second Bessel beams, respectively. (cf. Supplementary Section 3 for details.) If $(m^{(1)}, m^{(2)}) = (+\ell, -\ell)$, then ${}_{scat}F_\phi$ can be largely suppressed with the use of a TE ($C_E = 0$) or a TM ($C_H = 0$) mode because the third element of the matrix becomes smaller if either TE or TM mode is used :

$$\begin{bmatrix} (F_p)_\phi\big|_{(m^{(1)},m^{(2)})=(+l,-l)} \\ (F_m)_\phi\big|_{(m^{(1)},m^{(2)})=(+l,-l)} \\ (F_{pm})_\phi\big|_{(m^{(1)},m^{(2)})=(+l,-l)} \end{bmatrix}$$

$$\approx \begin{bmatrix} 0 \\ 0 \\ -\frac{1}{c} Re\left\{ a_{Mie}\, b_{Mie}^{*}\, \frac{1}{2} l \left(\cos 2\theta_{cone}^{(2)} - \cos 2\theta_{cone}^{(1)}\right)\frac{(\lambda/2\pi)}{\rho} |C|^2 J_l^2 \right\} \end{bmatrix} ; \left[ C = \begin{cases} C_E \text{ in TM mode} \\ C_H \text{ in TE mode} \end{cases} \right] \tag{5d}$$

This ensures a negligible azimuthal component of the scattering force compared to the naturally much stronger gradient force in the same direction.

$$(F_{pm})_\phi\big|_{(m^{(1)},m^{(2)})=(+l,-l)} \ll F_{grad}^{(\phi)}(max)\big|_{(m^{(1)},m^{(2)})=(+l,-l)} \tag{5e}$$

Superimposing two different (of helical nature) order Bessel beams (+$\ell$, -$\ell$), a '2$\ell$' number (cf. Eq.(5b)) of intensity pattern is created in each transverse plane, due to the superposition, (cf. Eq. (5a)), of the electric fields of the two beams. Even if the magnitudes of these fields of the beam pair are same, they have different and opposite helical phases. In their interference, (Fig. 7), the order sets the spacing of interference fringes.

On the other hand, while superimposing Bessel beams having the same order, the magnitude of superimposed electric fields just doubles and hence no intensity variation in the azimuthal direction can be created, (cf. Fig. 2s(a)), the intensity pattern will just follow the corresponding Bessel beam intensity profile. The use of different and reversed order Bessel beams is also responsible for multiple trap regions in the azimuthal plane (i.e. $\left|\mathrm{m}^{(1)}\right| + \left|\mathrm{m}^{(2)}\right| < 2 \times \max\left(\left|\mathrm{m}^{(1)}\right|, \left|\mathrm{m}^{(2)}\right|\right)$; where $\left|\mathrm{m}^{(1)}\right| \times \left|\mathrm{m}^{(2)}\right| < 0$). Based on these ideas, a completely new technique of manipulating multiple particles is demonstrated later in this article.

## D. THEORETICAL CONDITIONS FOR STABLE OPTICAL MANIPULATION

Fig. 4a shows how the particle moves as the minimum potential region shifts. Both the linear velocity and the angular speed of the particle, (cf. Fig. 4b in the main article and Fig. 4s in the supplement) have been derived to be, (cf. Supplementary Section 4.1 for the detailed proof):

$$v_{z.l}^{(H)} = v_{z.l}^{(L)} cos^2\left(\tan^{-1}\frac{m^{(1)} - m^{(2)}}{\beta_l^{(1)} - \beta_l^{(2)}}\right) = \frac{\left(\omega^{(1)} - \omega^{(2)}\right)}{(m^{(1)} - m^{(2)})^2 + \left(\beta_l^{(1)} - \beta_l^{(2)}\right)^2} \times \left(\beta_l^{(1)} - \beta_l^{(2)}\right) \approx v_{z.l}^{(L)} \tag{6}$$

$$\Omega_{\phi}^{(H)} = \Omega_{\phi}^{(C)} sin^2\left(\tan^{-1}\frac{m^{(1)} - m^{(2)}}{\beta_l^{(1)} - \beta_l^{(2)}}\right) = \frac{\left(\omega^{(1)} - \omega^{(2)}\right)}{(m^{(1)} - m^{(2)})^2 + \left(\beta_l^{(1)} - \beta_l^{(2)}\right)^2} \times \left(m^{(1)} - m^{(2)}\right) \approx 0 \tag{7}$$

By setting a certain frequency difference between the two fields of the beam pair, the speed of the particle can easily be set. Eqs.(6) and (7) are valid until the components of the gradient force in longitudinal and azimuthal directions counter balance the disturbing forces, (especially the viscous drag and the gravitational force) in their respective directions separately.The maximum permissible levitation velocity $\left|v_{z.l}^{(H)}\right|$of the particle is limited by the expression:

$$\left|{}_{grad}^{max}F_{z.l}\right| > \left|{}_{curl}F_{z.l} + {}_{scat}F_{z.l} + V\left(D_p - D_m\right)g\cos\angle t\right|_{max} + 6\pi r_p \vartheta_{liquid}\left|v_{z.l}^{(H)}\right| \tag{8}$$

Here $V$ is the volume of the particle, $D_p$ and $D_m$ are the densities of the particle and the medium respectively, $g$ is the gravitational acceleration. Eq.(6) is only applicable until the condition described by the inequality (8) holds. If (8) is not fulfilled because of an insufficient longitudinal gradient force, the azimuthal gradient force will make the particle rotate in the same transverse plane. But if the angular velocity of the particle is extremely high, the maximum possible binding force ${}_{grad}^{max}F_{\rho.l}$ in the radial direction may not be able to counter-balance the huge centrifugal force $F_c = m_p \left(\Omega_{\phi}^{(C)}\right)^2 r_{ring}$ , $m_p$ being the mass of the particle and $r_{ring}$ standing for the radial distance at any point on the elliptical ring (cf. Fig. 1s(b)). So, to ensure that the centrifugal force cannot make the radial binding vulnerable, the inequality given below must be maintained.

$$\left|{}_{grad}^{max}F_{\rho.l}\right| > \left|{}_{curl}F_{\rho.l} + {}_{scat}F_{\rho.l} + V\left(D_p - D_m\right)g \sin\angle t\right|_{max} + m_p\left(\Omega_{\phi}^{(C)}\right)^2 r_{ring} \tag{9}$$

Again, if the tangential velocity of the particle $v_{\psi.l}^{(p)} = \Omega_{\phi}^{(C)} r_{ring}$ (where $b \leq r_{ring} \leq a$) is too high to make the viscous force ${}_{visc}F_{\psi.l} = 6\pi r_p \vartheta_{liquid}\left(-v_{\psi.l}^{(p)}\right)$ large enough to exceed the maximum possible azimuthal binding force ${}_{grad}^{max}F_{\psi.l}$, the binding phenomena of the particle (with the moving intensity pattern) will be destroyed. So, the following inequality must be maintained for the sake of stable tangential binding:

$$\left|{}_{grad}^{max}F_{\psi.l}\right| > \left|{}_{curl}F_{\psi.l} + {}_{scat}F_{\psi.l} + V\left(D_p - D_m\right)g \sin\angle t\right|_{max} + 6\pi r_p \vartheta_{liquid}\left|\Omega_{\phi}^{(C)}\right| r_{ring} \tag{10}$$

The inequalities (9) and (10) provide the maximum possible angular speed that a particle can achieve. If both conditions given by Eqs.(8) and (10) fail, the particle will not be able to follow the moving potential either in the longitudinal or the azimuthal directions.

For example, a silicon-di-oxide particle having radius 100nm, completely immersed in water, can be successfully levitated from the liquid, (cf. Fig.5a), or successfully rotated in the liquid, (cf. Fig.5b), by using two second ordered Bessel beams, having an opposite helix from each other, at the wavelength of approximately 1 $\mu m$(Cf. Video File 2 : Part 6 and Part 9).

**Table 1: Gas, Particle and Liquid Properties**

| Gas Medium | Particle | Liquid Medium |
|---|---|---|
| $\eta_g = 1$ | $\eta_p = 1.535$at $\lambda = 1\ \mu m$ | $\eta_m = 1.33$ |
| $\rho_g = \rho_{air}$ | $\rho_p = 2650\ kgm^{-3}$ | $\rho_m = 1000\ kgm^{-3}$ |

For a 15 $kHz$ frequency-difference [55-62], an electric field of magnitude $7 \times 10^6\ Vm^{-1}$ and cone-angles $40^0$and $65^0$of the superimposed beams, a successful linear movement along the beam-axis occurs at a speed of 3.28 $cms^{-1}$, (cf. Fig.5a. See also Tables 2(a) and 3(a)). The longitudinal viscous drag on the particle for this speed is approximately $F_{visc}^{(z)} = -62\ pN$ and the gradient force in this direction is strong enough ($-107.89\ pN \leq F_{grad}^{(z)} \leq +107.89\ pN$) to balance the dragging force from the opposite direction ($F_{grad}^{(z)} = +62\ pN$) so that the particle experiences no net force, (cf. Table 4(a)). The maximum intensity [63-68] is in moderate level.

**Table 2(a): Properties of theBessel beam-pair**

| Parameter | Bessel Beam 1 | Bessel Beam 2 |
|---|---|---|
| $(C_E, C_H)$ | $(7 \times 10^6\ Vm^{-1}, 0\ Am^{-1})$ | $(7 \times 10^6\ Vm^{-1}, 0\ Am^{-1})$ |
| $m$ | $+2$ | $-2$ |
| $\omega$ | $\omega^{(1)} = 2\pi \times 300\ TeraHz$ | $\omega^{(2)} = \omega^{(1)} \pm (2\pi \times 15\ kHz)$ |
| $\theta_{cone}$ | $40^0$ | $65^0$ |
| $\beta_{gas}$ | $\beta_{gas}^{(1)} = \frac{\omega^{(1)}}{c} \cos\theta_{cone}^{(1)}$ | $\beta_{gas}^{(2)} = \frac{\omega^{(2)}}{c} \cos\theta_{cone}^{(2)}$ |

**Table 3(a): Beam, Particle and Liquid Properties**

| Parameter | Value |
|---|---|
| Incidence Angle | $45^0$ |
| Particle Radius | $100\ nm$ |
| Liquid (water) Viscosity | $1.002 \times 10^{-3}\ Pa\ s$ |
| Longitudinal velocity of the particle | $\sim 3.28\ cm\ s^{-1}$ |
| Azimuthal speed of the particle | $1.5 \times 10^{-9} rad\ s^{-1}$ |
| Maximum Intensity on the Particle | $\sim 4.5343 \times 10^9\ W\ cm^{-2}$ |

**Table 4(a): Gradient, Viscous/Centrifugal and Gravitational Forces in different direction**

| Direction | Gradient Force | Viscous/Centrifugal Force (opposite to Gradient Force) | Gravitational Force |
|---|---|---|---|
| Longitudinal | $-107.89\ pN \leq F_{grad}^{(z)} \leq +107.89\ pN$ | $\left\|F_{visc}^{(z)}\right\| = 62\ pN$ | $F_{grvt}^{(z)} = 5.74 \times 10^{-5}\ pN$ |
| Azimuthal | $-365.18\ pN \leq F_{grad}^{(\psi)} \leq +365.18\ pN$ | $\left\|F_{visc}^{(\psi)}\right\| = 6.1 \times 10^{-11}\ pN$ | $F_{grvt}^{(\psi)} = 3.60 \times 10^{-5}\ pN$ |
| Radial | $F_{grad}^{(\rho)} \leq +309.37\ pN$ | $\left\|F_{cent}^{(\rho)}\right\| = 1.6 \times 10^{-26}\ pN$ | $F_{grvt}^{(\rho)} = 3.60 \times 10^{-5}\ pN$ |

Also, in case of a 15 $kHz$ frequency-difference, electric field of magnitude $4 \times 10^7\ Vm^{-1}$ and cone-angles $45^0$ and $50^0$ of the superimposed beams, a successful rotation around the beam-axis occurs at $3750\ rev\ s^{-1}$ (cf.Fig.5b and Tables 2(b) and 3(b)). The azimuthal viscous drag on the particle for this rotational speed is approximately $F_{visc}^{(\psi)} = -35.8\ pN$ and the gradient force in this direction is strong enough ($-139.88\ pN \le F_{grad}^{(\psi)} \le +139.88\ pN$)to balance the dragging force from the opposite direction ($F_{grad}^{(\psi)} = +35.8\ pN$)so that the particle experiences no net force, (cf. Table 4(b)). In this case, no levitation occurs because of weak longitudinal gradient force$\left(\left|F_{grad}^{(z)}\right| < 173\ pN\right)$.

**Table 2(b): Properties of the Bessel beam-pair**

| Parameter | Bessel Beam 1 | Bessel Beam 2 |
|---|---|---|
| $(C_E, C_H)$ | $(4 \times 10^7\ Vm^{-1}, 0\ Am^{-1})$ | $(4 \times 10^7\ Vm^{-1}, 0\ Am^{-1})$ |
| $m$ | $+2$ | $-2$ |
| $\omega$ | $\omega^{(1)} = 2\pi \times 300\ TeraHz$ | $\omega^{(2)} = \omega^{(1)} \pm (2\pi \times 15\ kHz)$ |
| $\theta_{cone}$ | $40^0$ | $50^0$ |
| $\beta_{gas}$ | $\beta_{gas}^{(1)} = \left(\omega^{(1)}/c\right) \cos\theta_{cone}^{(1)}$ | $\beta_{gas}^{(2)} = \left(\omega^{(2)}/c\right) \cos\theta_{cone}^{(2)}$ |

**Table 3(b): Beam, Particle and Liquid Properties**

| Parameter | Value |
|---|---|
| Incidence Angle | $45^0$ |
| Particle Radius | $100\ nm$ |
| Liquid (water) Viscosity | $1.002 \times 10^{-3}\ Pa\ s$ |
| Longitudinal velocity of the particle | $0\ cm\ s^{-1}$ |
| Azimuthal speed of the particle | $3750\ rev\ s^{-1}$ |
| Maximum Intensity on the Particle | $\sim 1.5 \times 10^9\ W\ cm^{-2}$ |

**Table 4(b): Gradient, Viscous/centrifugal and Gravitational Forces in different direction**

| Direction | Gradient Force | Viscous/Centrifugal Force | Gravitational Force |
|---|---|---|---|
| Longitudinal | $-10.92\ pN \le F_{grad}^{(z)} \le +10.92\ pN$ | $\left\|F_{visc}^{(z)}\right\| = 0\ pN$ | $F_{grvt}^{(z)} = 5.74 \times 10^{-5}\ pN$ |
| Azimuthal | $-139.88\ pN \le F_{grad}^{(\psi)} \le +139.88\ pN$ | $\left\|F_{visc}^{(\psi)}\right\| = 35.8\ pN$ | $F_{grvt}^{(\psi)} = 3.60 \times 10^{-5}\ pN$ |
| Radial | $F_{grad}^{(\rho)} \le +62.71\ pN$ | $\left\|F_{cent}^{(\rho)}\right\| = 0.005\ pN$ | $F_{grvt}^{(\rho)} = 3.60 \times 10^{-5}\ pN$ |

This pair of Bessel beams can also be used to trap a particle at a certain location, (cf. Fig.6). Refs.[69-71], regarding laser pulse duration, prove that switching laser light in the micro-second range interval is possible in practical cases (Cf. Video File 2 : Part 10).

## E. OPTICAL MANIPULATION OF MULTIPLE PARTICLES BY SELECTING BESSEL ORDERS

One of the major findings of the present work is the theoretical foundation of multiple particle manipulation inside a liquid medium, (cf. Fig. 8a of the main article and Fig. 5s of the Supplement). Necessary physical explanations behind the increasing of the trapping regions for multiple particle translation or rotation, (because of different and reversed order Bessel beams), are given here, (cf. Fig.7). For multiple particle manipulation purposes [4], a selection of the orders of the Bessel beam pair of mutually reversed helical nature, is of great importance. If the absolute values of the reverse orders are not exactly equal, the azimuthal scattering force in the clockwise direction (cf. Eq.(26s)) may not fully balance (or be balanced by) the azimuthal scattering force component in the counter-clockwise direction, (cf. Eq.(27s)). But this resultant non-zero azimuthal scattering force is expected to be much smaller than the azimuthal gradient force and hence may cause no significant disturbance in optical manipulation.

## F. OPTICAL MANIPULATION ACCORDING TO PARTICLE-SIZE

As the size of the particles varies, the whole beam-intensity system can be made compatible simply by keeping the size-parameter unchanged, (cf. Fig. 8b). If the radius of the new particle is $N$ times the radius of the previous particle: $r_{new}^{(part)} = N r_{old}^{(part)}$, then the new wavelength should be exactly $N$ times that of the previous one: $\lambda_{new} = N\lambda_{old}$, (cf. Supplementary Section 6 for details). On maintain the ratio of the new and previous intensities as $I_{new} : I_{old} = \left[\frac{\eta_m}{2\,c}\left(3\,\frac{\eta_p^2-\eta_m^2}{\eta_p^2+2\eta_m^2}\right)\right]_{\lambda_{old}} : \left[\frac{\eta_m}{2\,c}\left(3\,\frac{\eta_p^2-\eta_m^2}{\eta_p^2+2\eta_m^2}\right)\right]_{\lambda_{new}}$, the following relationships can be achieved:

$$\begin{bmatrix} {}^{new}_{grad}F_{z.l} \\ {}^{new}_{grad}F_{\rho.l} \\ {}^{new}_{grad}F_{\phi.l} \end{bmatrix} = N^2 \begin{bmatrix} {}^{old}_{grad}F_{z.l} \\ {}^{old}_{grad}F_{\rho.l} \\ {}^{old}_{grad}F_{\phi.l} \end{bmatrix} \tag{11}$$

The disturbing forces are:

$$\begin{bmatrix} F_{new}^{(visc.Z)} \\ F_{new}^{(cent.\rho)} \\ F_{new}^{(visc.\psi)} \end{bmatrix} = \begin{bmatrix} N^2 F_{old}^{(visc.Z)} \\ N^4 F_{old}^{(cent.\rho)} \\ N^2 F_{old}^{(visc.\psi)} \end{bmatrix} \tag{12}$$

From Eqs.(11) and (12) we see that by letting $\lambda_{new} = N\lambda_{old}$ the binding force proportionally increases as $N^2$, whereas two of the three disturbing forces increase proportionally with $N^2$and one with $N^4$.As the unchanged size-parameter offers the relationships $_{grad}^{new}F_{z.l} : F_{new}^{(visc.Z)} = {}_{grad}^{old}F_{z.l} : F_{old}^{(visc.Z)}$ and $_{grad}^{new}F_{\phi.l} : F_{new}^{(visc.\psi)} = {}_{grad}^{old}F_{\phi.l} : F_{old}^{(visc.\psi)}$, a stable optical manipulation can be easily maintained on any size of particles. Although $_{grad}^{new}F_{\rho.l} : F_{new}^{(cent.\rho)} \neq {}_{grad}^{old}F_{\rho.l} : F_{old}^{(cent.\rho)}$, the centrifugal force being much weaker than the radial gradient force (cf. Table 4) should not cause any significant disturbance.

As an example, the use of $\sim 1\mu m$ wavelength is adequate to successfully levitate particles of radius not exceeding $100\,nm$. For particles of radius around $1\mu m$, beams with larger wavelengths $\sim 10\mu m$ are perfect for levitation purposes. Thus, for a given wavelength, there exists a maximum size of particles above which levitation is not possible. Hence to increase this maximum permissible particle size, the wavelength should be enlarged. This wavelength control, which underlines the optical sorting of particles [4] of different sizes, is demonstrated in Supplementary Section 6 (Cf. Video File 2: Part 7).

## G. OPTICAL MANIPULATION WITH NON-COAXIAL BESSEL BEAMS

The use of a co-axial beam pair is not a limitation of our work because in case of non-coaxial beam pair, levitation is also possible, even though within a certain range. If the beams are incident from air to liquid with incidence angles $\angle i_1$ and $\angle i_2$ respectively, as explained in Fig. 1(c), then the length of levitation range is given by $d_{range} = 2R_2\zeta_2/\sin\angle\xi_2 = 2R_1\zeta_1/\sin\angle\xi_1$ (cf. Fig. 9(a)) where $\angle\xi_1$ and $\angle\xi_2$ are the angles made by the particle levitation direction with the transmitted beam-axis of first and second beams, respectively; and the geometric parameters $\zeta_1$ and $\zeta_2$ depend on the incidence angles (cf. Supplementary Section 7 for details) :

$$\begin{bmatrix} \angle\xi_1 \\ \angle\xi_2 \end{bmatrix} = \begin{bmatrix} \tan^{-1}\dfrac{\sin(\angle t_1 - \angle t_2)}{(R_2\zeta_2)/(R_1\zeta_1) + \cos(\angle t_1 - \angle t_2)} \\ (\angle t_1 - \angle t_2) - \tan^{-1}\dfrac{\sin(\angle t_1 - \angle t_2)}{(R_2\zeta_2)/(R_1\zeta_1) + \cos(\angle t_1 - \angle t_2)} \end{bmatrix} \tag{13a}$$

where,

$$\begin{bmatrix} \angle t_1 & \zeta_1 \\ \angle t_2 & \zeta_2 \end{bmatrix} = \begin{bmatrix} \sin^{-1}(\eta_G/\eta_L)\sin\angle i_1 & \sqrt{1 + [1 - (\eta_G/\eta_L)^2]\tan^2\angle i_1} \\ \sin^{-1}(\eta_G/\eta_L)\sin\angle i_2 & \sqrt{1 + [1 - (\eta_G/\eta_L)^2]\tan^2\angle i_2} \end{bmatrix} \tag{13b}$$

The velocity of the particle is then given by $v_{z.l}^{(p)} \approx (\omega^{(1)} - \omega^{(2)}) / (\beta^{(1)} \cos\angle\xi_1 - \beta^{(2)} \cos\angle\xi_2)$. One can use the relation between the maximum possible levitation range and the angle between two beam axes as a possible mechanism to levitate particles from a specific depth. The larger is the angle between two beam axes, the shorter the levitation range becomes (Cf. Video File 2: Part 8). Fig. 9(b) demonstrates particle levitation with time by projecting two Bessel beams of reverse order at incident angles: $\angle i^{(1)} = +5^0$ and $\angle i^{(2)} = -5^0$on the water surface. The relation between the maximum possible depth, from which the particle can be safely levitated, and the incidence angles of the two Bessel beams, is demonstrated in Fig. 10. An angle difference between the two beams can serve as a new degree of freedom to trap in three dimensions, or to rotate, or to levitate fully immersed particles from specific depths.

## CONCLUSION

A persistent challenge with current opto-fluidic devices is achieving accurate manipulation of trapped particles in a controlled way [72]. Our proposed light source, and its advantage in controlled levitation of particles from liquid media, should lead to a fully computer-based automatic solution [73] of sorting, detecting, transporting, and isolating specific and desired particles, or molecules, without destruction [21]. If fully immersed dielectric particles are Rayleigh ones, (i.e. of size of e.g. virus or bacteria between 20 and 200 nm [74]), or different to Mie particles [38], optical pulling of such multiple particles has not been successful so far. We have proposed in this paper not only the resolution to such problems, but also a way that may be pioneering for almost all kinds of controlled manipulations, (i.e. rotation, lateral, and azimuthal movements) for fully immersed multiple objects. This is done by increasing the number of binding regions with a single optical setup. We believe that our work will have an impact on various real world applications such as to extract unwanted biological cells [75], control atmospheric pollution, fractionate blood particles [43], cell signaling research [44], controlled manipulation of DNA and live cells [25, 45], in an innocuous manner.

## ACKNOWLEDGMENT

We acknowledge discussions with Prof. Qiu Cheng Wei from NUS, as well as with Prof. Asif Zaman and Prof. AnisuzzamanTalukder from BUET. MN-V is supported by the Spanish MINECO through FIS2012-36113-C03-03 and FIS2014-55563-REDC research grants.

**Video File 1 and 2:** https://sites.google.com/site/ayedalsayem1/research-interest/publications

**Supplementary article:** Supplementary online article contains detail derivations, several simulation results and the details of the numerical calculations.

## Figure and Caption List

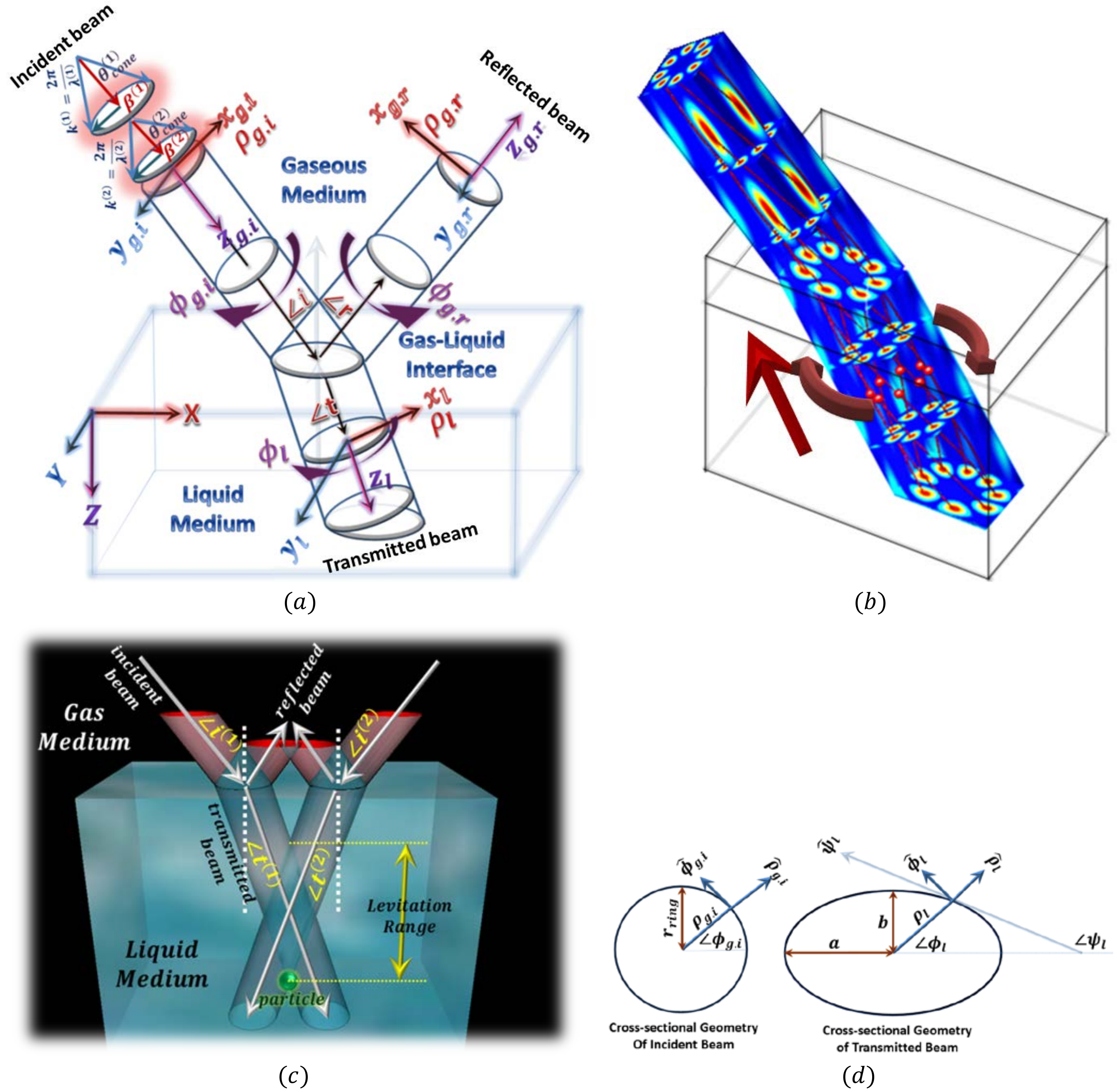

Fig.1: Continuous and three dimensional stable manipulations of fully immersed particles by using optical gradient force generated by the super-imposed co-axial Bessel beam-pair of reverse orders and different frequencies. (a) The co-ordinate system of the co-axial and superimposed beam-pair having different cone-angles and different wave-numbers. Between the two portions of incident beam namely transmitted and reflected beam, first one (transmitted beam in the liquid medium) is the main concern in this work.

The beam-pair must have different paraxial angles $\left(\theta_{cone}^{(1)} \neq \theta_{cone}^{(2)}\right)$ to maintain the change of intensity in the propagation direction, different orders $\left(m^{(1)} \neq m^{(2)}\right)$ to create azimuthal change of intensity pattern and different frequencies of $\left(\omega^{(1)} \neq \omega^{(2)}\right)$ for particle movement. (b) Ultimate intensity profile of the incident and transmitted beam-pair in gas-medium and liquid medium respectively are shown, which is perfectly capable of levitating or rotating a group of dielectric particles of any radius (for which equation (2) is applicable) located in liquid medium. The linear or rotational speed can be varied from very low value to high value easily by overcoming the phenomena of the disturbing forces like gravitational pull, viscous drag, centrifugal force and relatively much weaker scattering. This system is also capable of trapping a particle at any desired location inside the liquid. (c) Non coaxial setup for tractor beam: two Bessel beams of different helical nature are incident from air to liquid medium with an angle $\angle i_1$ and $\angle i_2$ respectively in such a way that the beam-axes intersect each other inside the liquid with an angle of $|\angle t_1| + |\angle t_2|$ (if beams are inclined at reverse directions) or $|\angle t_1| - |\angle t_2|$ (if beams are inclined at the same direction). Only the particles located within a certain range can be levitated using this beam set-up. (d) The circular and elliptical locus of equipotential region in the cross-section perpendicular to the incident (left) and transmitted (right) beam axis. The transmitted beam property inside water is not fully same as that of the incident beam. A non-parxial single gradientless Bessel beam may not be capable of pulling a dipole object by overcoming the viscous and other disturbing forces.

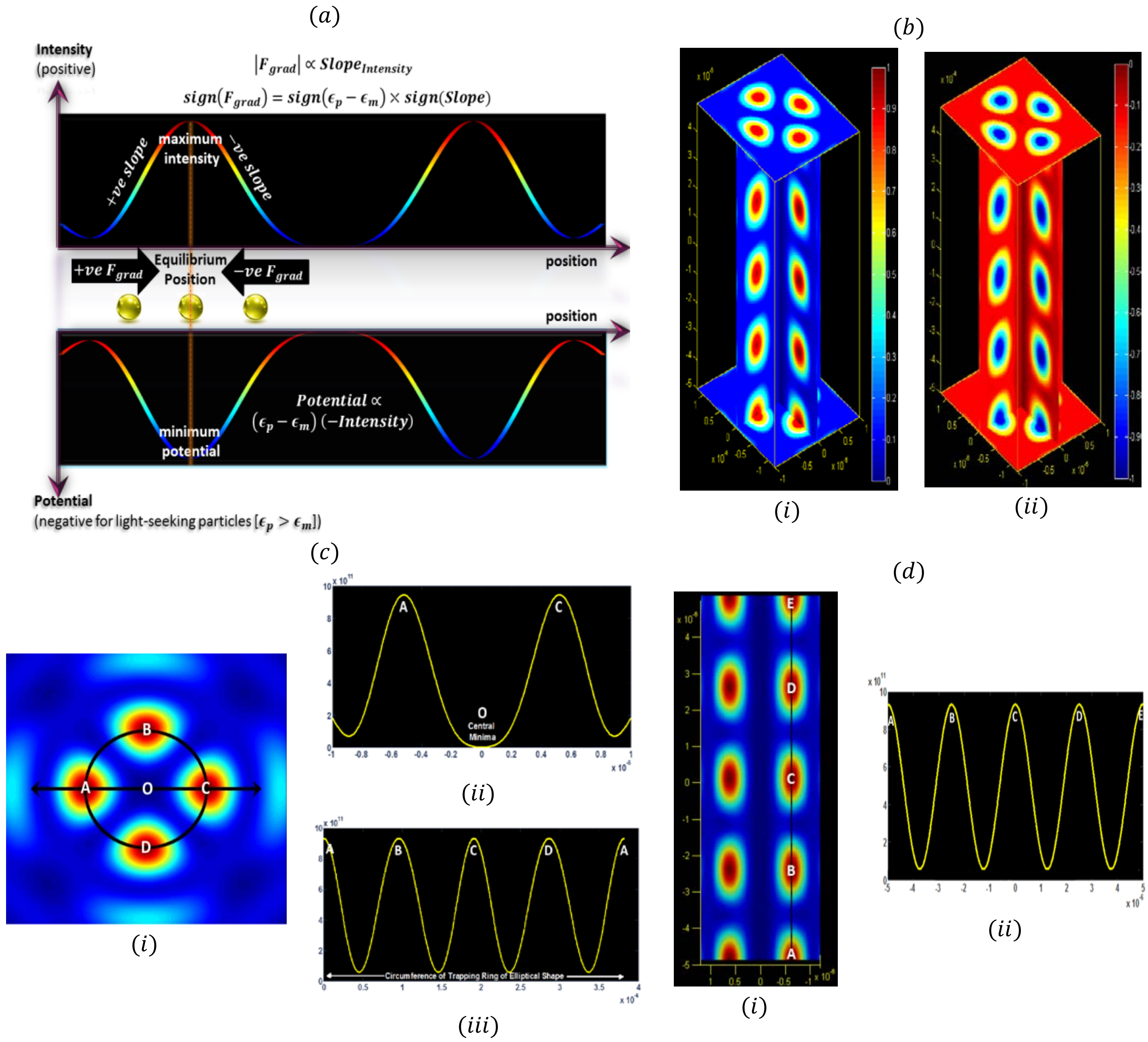

Fig.2: Optical binding of a dielectric particle at the point (equilibrium position) where net force is zero or in other words: the potential felt by the particle is the minimum. The beam parameters of the first and second beam are

$\left(m^{(1)}, \beta_{gas}^{(1)}, \theta_{cone}^{(1)}, C_E^{(1)}, C_H^{(1)}\right) = \left(+2, \sqrt{2}\pi \times 10^6\ radm^{-1}, \cos^{-1}(1/\sqrt{2}), 10^6 Vm^{-1}, 0\ Am^{-1}\right)$ and

$\left(m^{(2)}, \beta_{gas}^{(2)}, \theta_{cone}^{(2)}, C_E^{(2)}, C_H^{(2)}\right) = \left(-2, (2\sqrt{2}/5)\pi \times 10^6\ rad\ m^{-1}, \cos^{-1}(\sqrt{2}/5), 10^6 Vm^{-1}, 0\ Am^{-1}\right)$

respectively. The superimposed beam-pair is incident at the gas-liquid interface at an angle of $\angle i = 45^0$. (a) If the dielectric constant of the particle is greater than the dielectric constant of the medium, the direction of the gradient force is same as the sign of intensity-slope. Such a particle (shown in the figure) will be stabilized at the nearest maxima of intensity pattern i.e. at a position where it feels minimum potential. (b) The cross-sectional view of (i) Intensity-pattern (ii) Potential-profile (per unit volume of the particle) faced by a light seeking dielectric particle ($\eta_p > \eta_m$). (c) Extraction of one dimensional change of intensity pattern in azimuthal and radial direction separately from two dimensional simultaneous changes: (i) Transverse intensity pattern at $\rho$-$\phi$ plane. (ii) Radial intensity variation along the open-line AOC passing through the center O and any of the most intensified points. (iii)Azimuthal intensity variation along the closed-ring ABCDA passing through the most intensified points.(d) Longitudinal intensity variation along a line parallel to the beam-axis and passing through any of the four most intensified points at the transverse plane shown in (i) z$\rho$ plane and in (ii) graphical demonstration of intensity change along the open-line ABCDE in longitudinal direction.

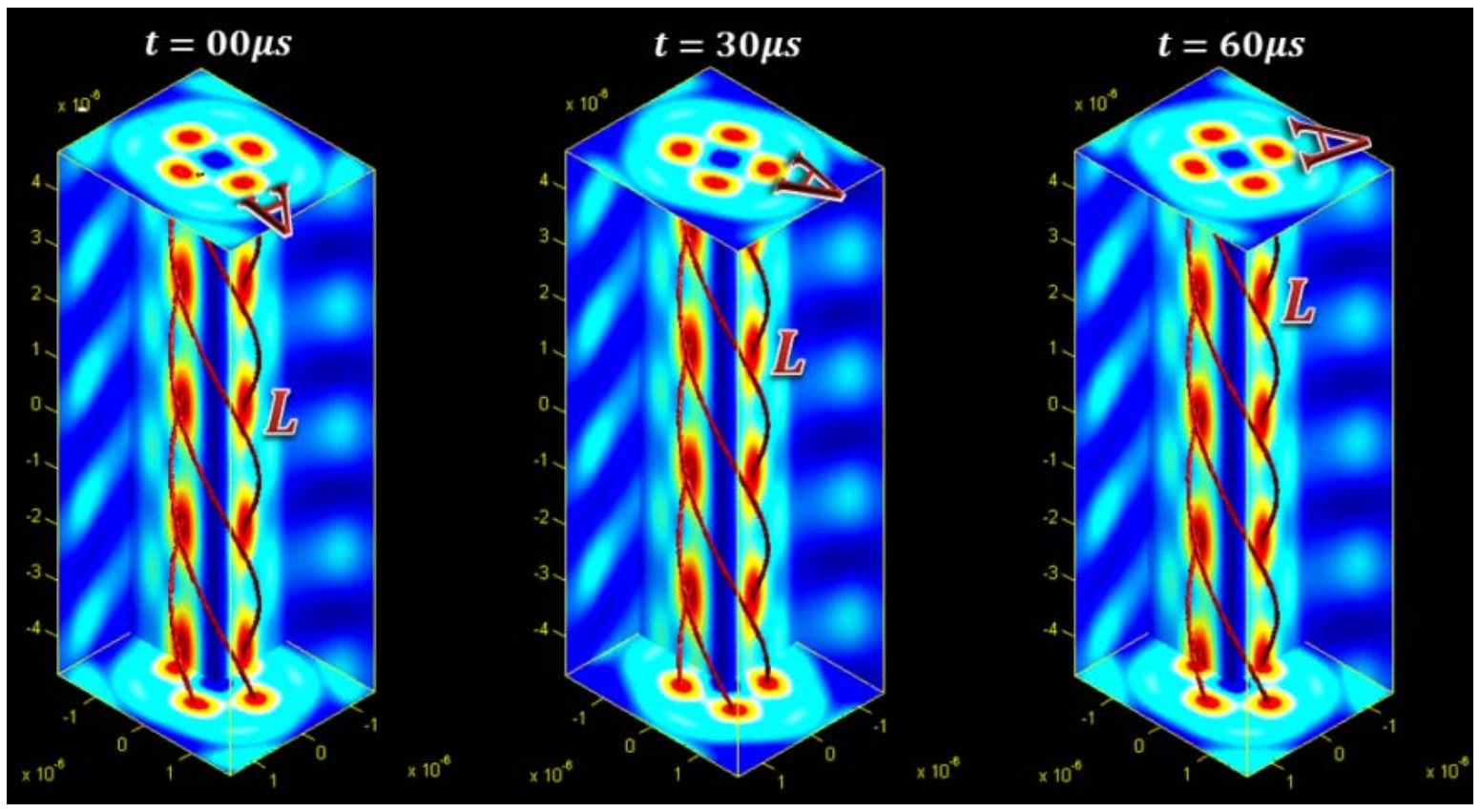

$(a)$

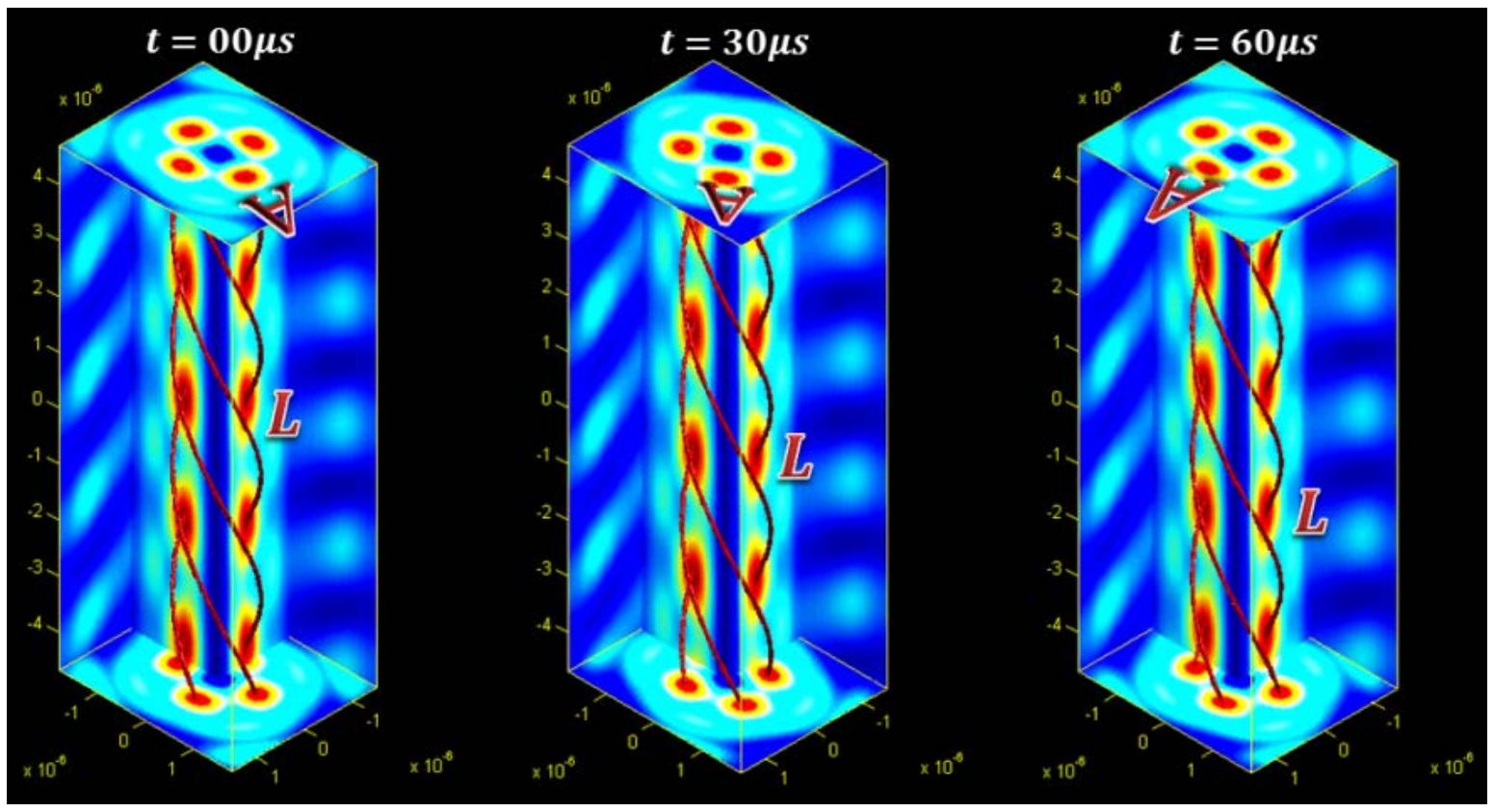

$(b)$

Fig. 3: An illustration of the movement of the intensity pattern (not the movement of particle) of the superimposed co-axial beam with time. It is important to note that the velocity of the intensity pattern

movement is not the velocity of the particle. Velocity of the particle will be discussed in next figure (Fig. 4).Here we have demonstrated the time domain simulation of intensity pattern or potential profile for $\omega^{(1)} \neq \omega^{(2)}$ where the beam parameters are $\left(m^{(1)}, \theta_{cone}^{(1)}, \omega^{(1)}, C_E^{(1)}, C_H^{(1)}\right) = (+2, 40^0, 2\pi \times 300\ THz, 10^6 Vm^{-1}, 0\ Am^{-1})$ and $\left(m^{(2)}, \theta_{cone}^{(2)}, \omega^{(2)}, C_E^{(2)}, C_H^{(2)}\right) = \left(-2, 65^0, \omega^{(1)} \pm 2\pi \times 15\ kHz, 10^6 Vm^{-1}, 0\ Am^{-1}\right)$. (a) The variation of intensity profile for $\omega^{(2)} < \omega^{(1)}$ in 3D view for $\omega^{(1)} - \omega^{(2)} = 2\pi \times 15\ kHz$. The change of intensity pattern at a certain cross-section accommodating the beam-axis when the pattern is changing at a the rate of $3.28175\ \mu m\ \mu s^{-1}$ in upward direction and the change of intensity pattern at a certain cross-section perpendicular to the beam axis when the pattern is changing at the rate of $3.75\ rev\ ms^{-1}$ in anti-clockwise (top-view) direction (b) Reverse directional change of intensity pattern obtained for $\omega^{(2)} > \omega^{(1)}$ by keeping $\omega^{(2)} - \omega^{(1)} = 2\pi \times 15\ kHz$ . The change of intensity pattern at a certain cross-section accommodating the beam axis when the pattern is changing at the rate of $3.28175\ \mu m\ \mu s^{-1}$ in downward direction and the change of intensity pattern at a certain cross-section perpendicular to the beam axis when the pattern is changing at the rate of $3.75\ rev\ ms^{-1}$ in clockwise (top-view) direction.

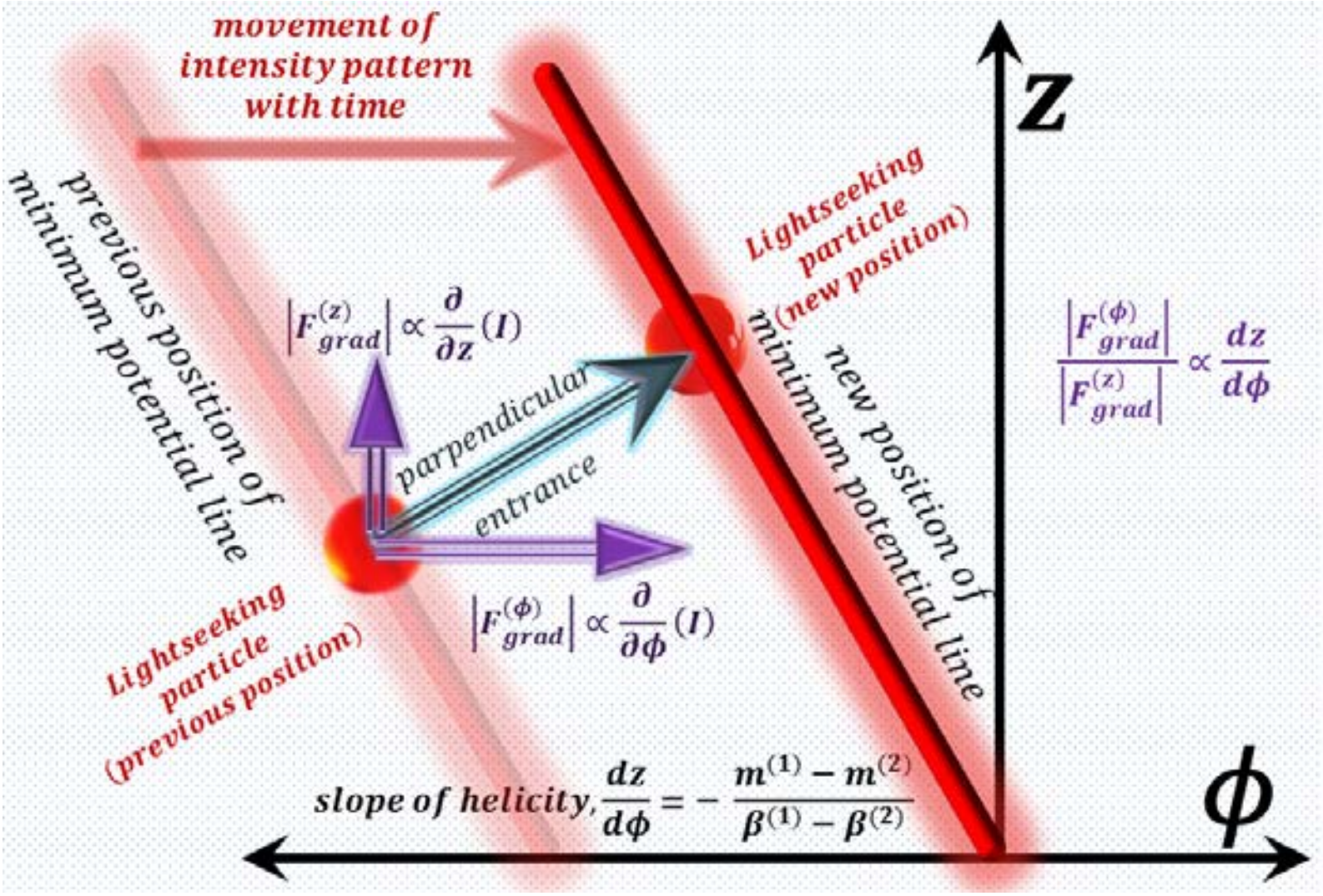

$(a)$

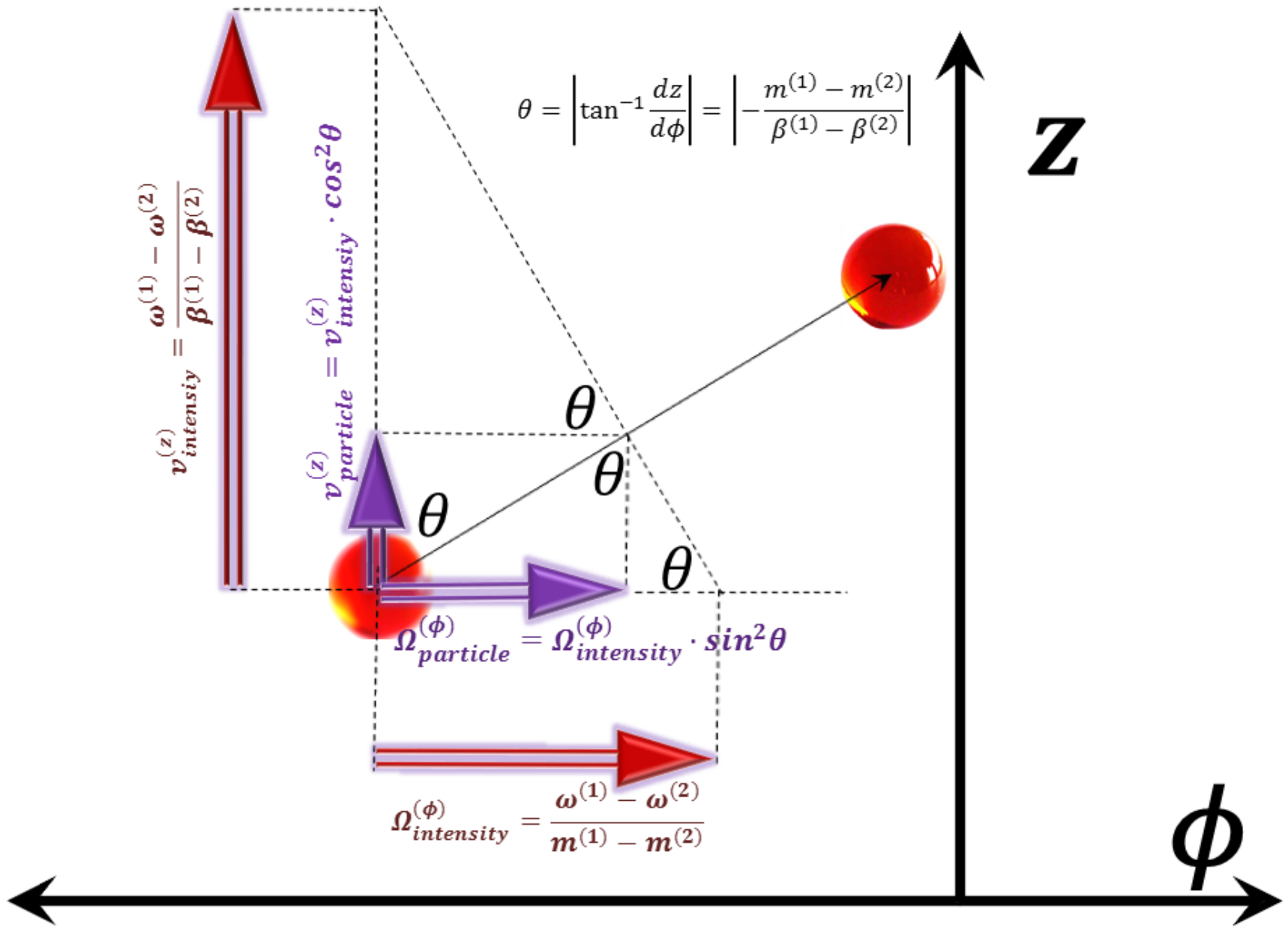

$(b)$

Fig.4: An illustration of the dynamic movement of the particle due to dynamic change of the intensity pattern. However, the velocity of the intensity pattern (discussed in previous figure) is not the velocity of the particle. (a) The particle travels through the shortest possible path to reach the relocated minimum potential region by dint of optical gradient force because the ratio of azimuthal component to longitudinal component of the gradient force is directly proportional to the slope of helicity of the intensity pattern. (b) Graphical explanation of the velocity of the particle following the intensity pattern (cf. supplementary section 4.1 for detailed mathematical insight).

$(a)$

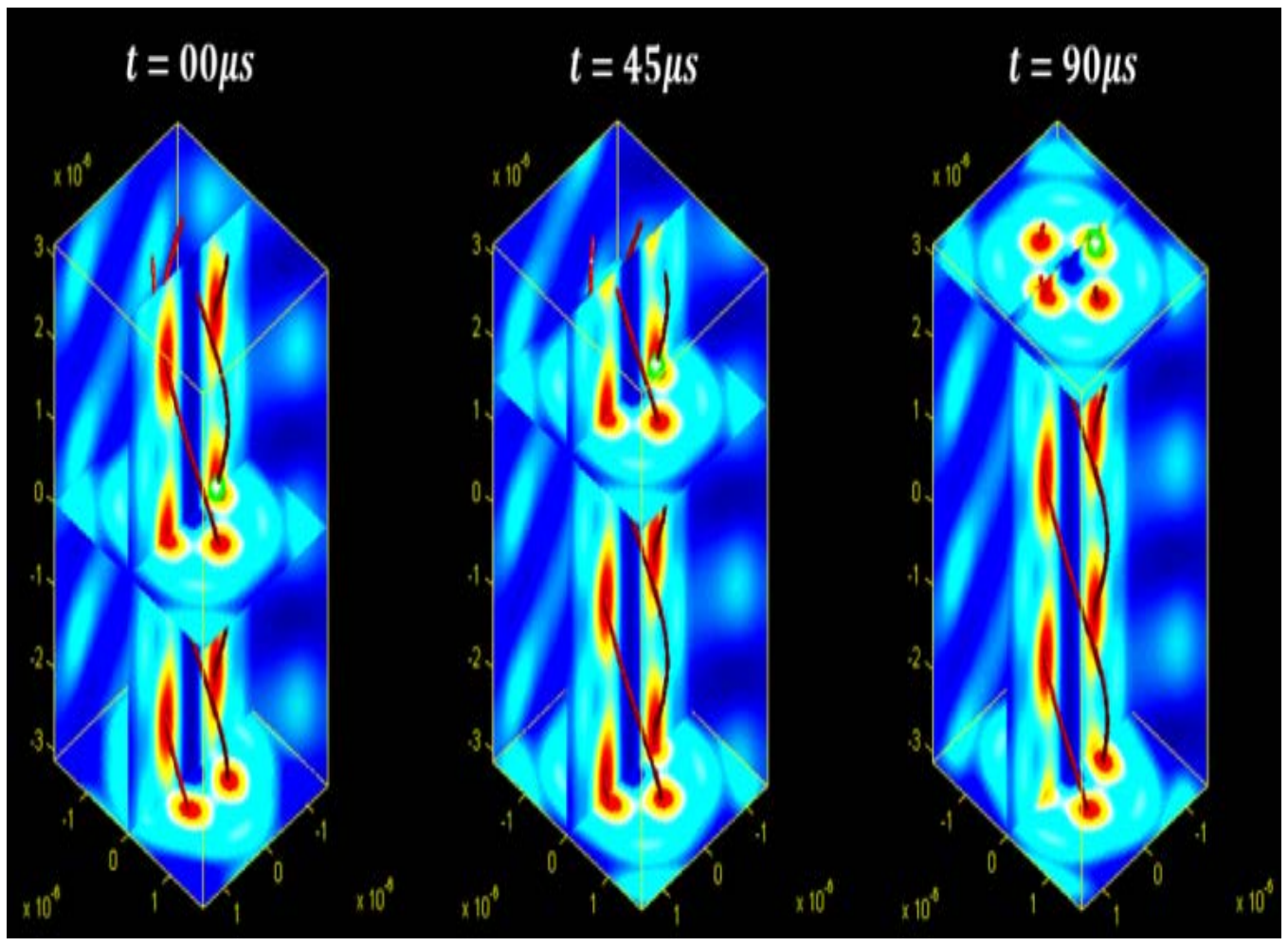

$(i)$

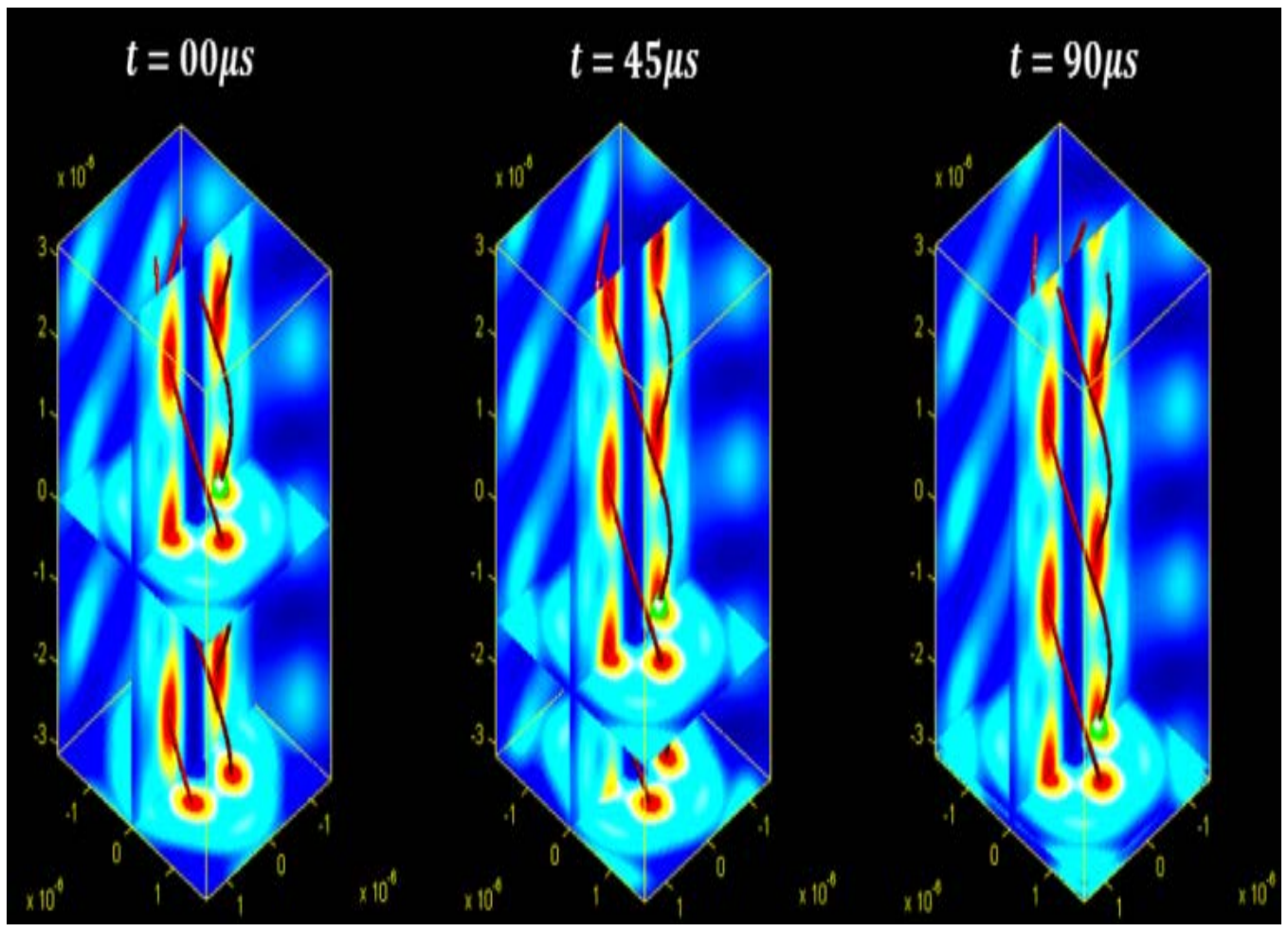

$(ii)$

$(b)$

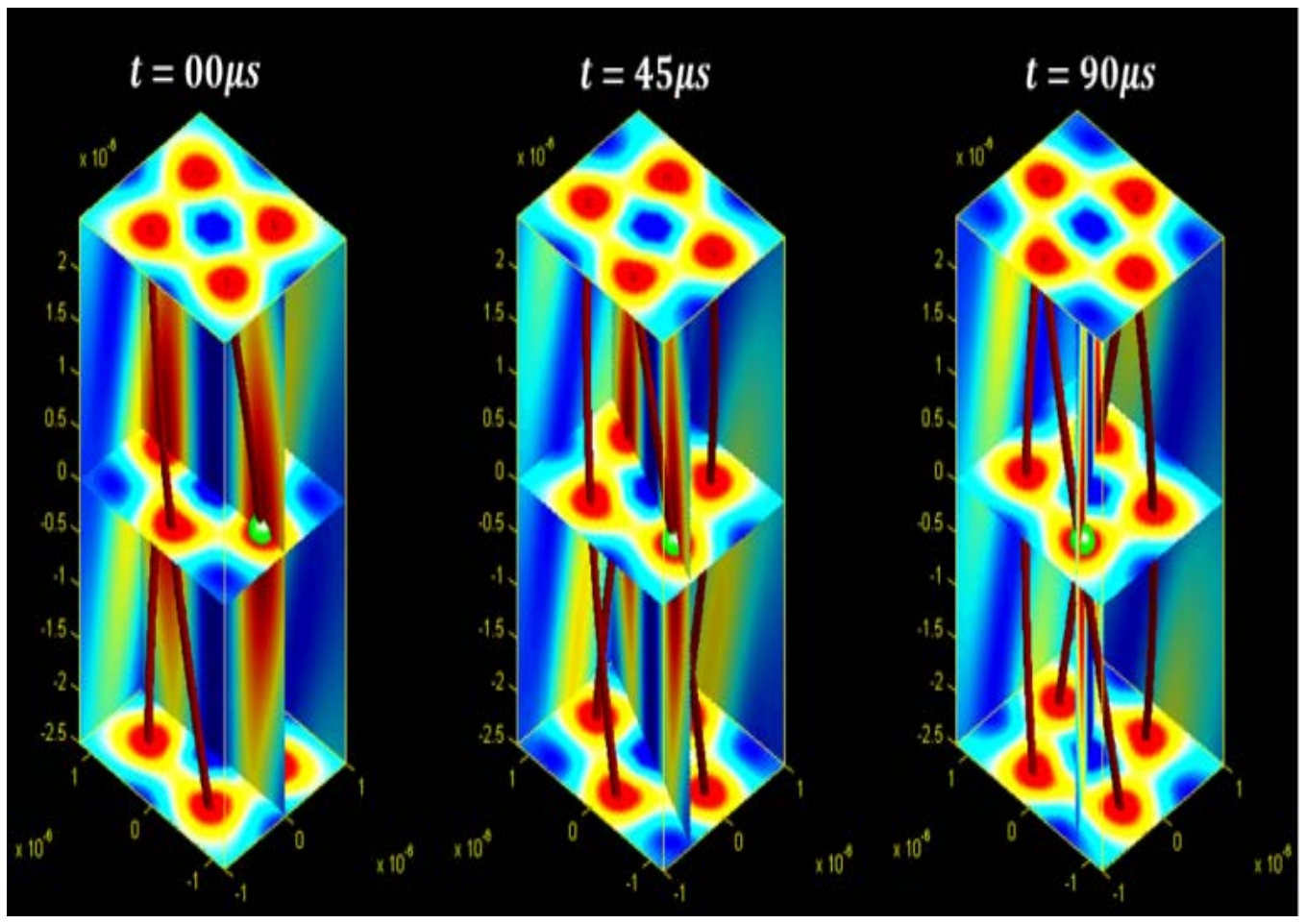

$(i)$

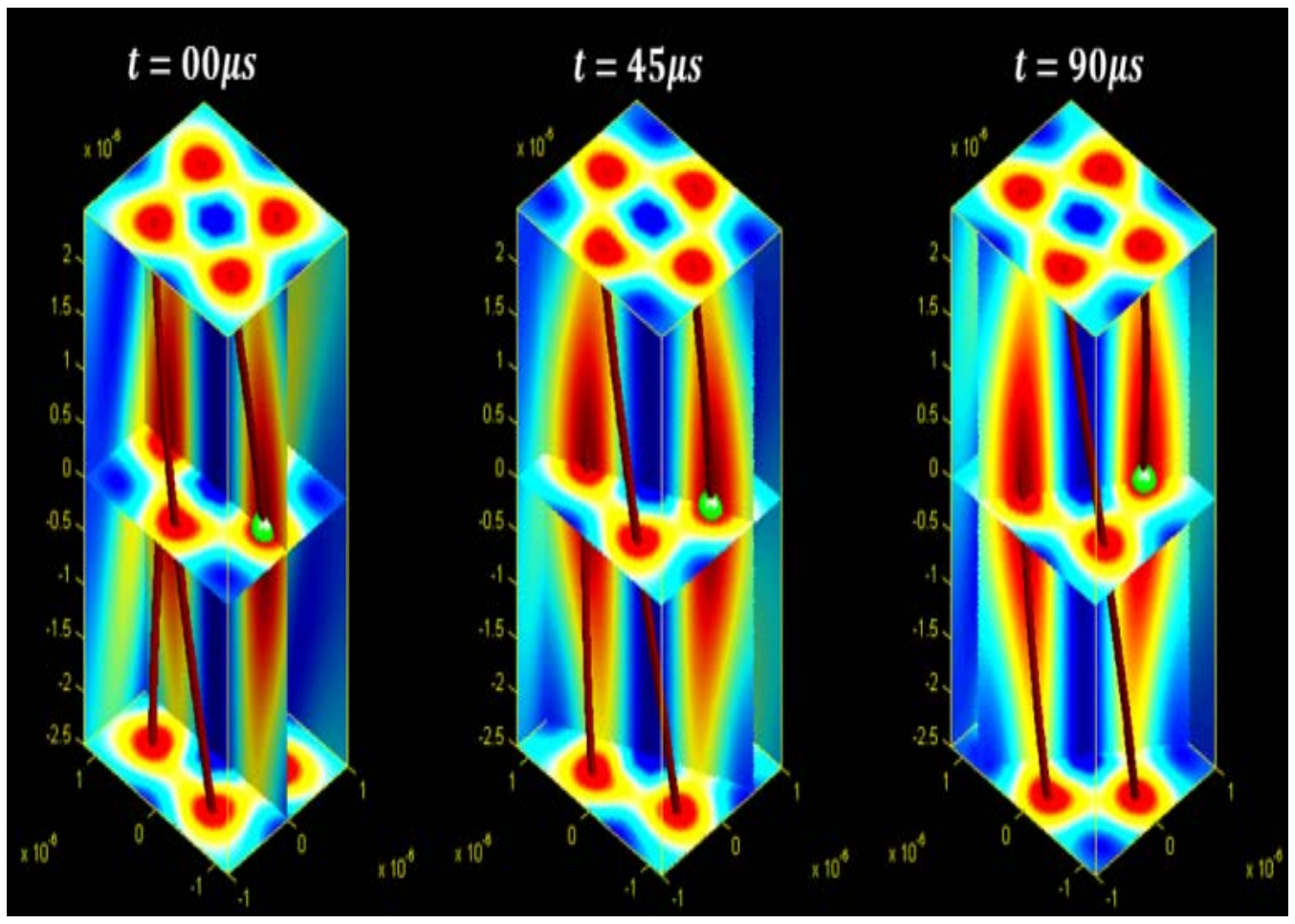

$(ii)$

Fig.5: In the above simulation the beams have the following parameters $\left(m^{(1)}, \theta^{(1)}_{cone}, \omega^{(1)}, C^{(1)}_H\right) = (+2, 40^0, 2\pi \times 300\ THz, 0\ Am^{-1})$&$\left(m^{(2)}, \theta^{(2)}_{cone}, \omega^{(2)}, C^{(2)}_H\right) = \left(-2, 65^0, \omega^{(1)} \pm 2\pi \times 15\ kHz, 0\ Am^{-1}\right)$.The wavelength used in the simulationis $1\mu m$. The density of the particle ($SiO_2$) is $2650\ kgm^{-3}$ and the refractive index at this wavelength is $1.535$.(a) If the paraxial angles are chosen as $\left(\theta^{(1)}_{cone}, \theta^{(2)}_{cone}\right) = (40^0, 65^0)$and $\left(C^{(1)}_E, C^{(2)}_E\right) = (7 \times 10^7 Vm^{-1}, 7 \times 10^7 Vm^{-1})$, longitudinal shift of a dielectric particle (green colored sphere) in either upward (i) or downward (ii) direction at a speed of approximately $3.284\ cm\ s^{-1}$with negligible angular displacement $(1.5 \times 10^{-9}\ rad\ s^{-1})$. The radius of the particle should be less than $0.15\mu m$ and during levitation of multiple particles, the inter-distance between nearest

particles in the transverse plane is approximately $0.832\mu m$. The maximum intensity faced by the particle is approximately $4.53 \times 10^9\ W\ cm^{-2}$ (b) In case of paraxial angles $\left(\theta_{cone}^{(1)}, \theta_{cone}^{(2)}\right) = (40^0, 50^0)$ and $\left(C_E^{(1)}, C_E^{(2)}\right) = (4 \times 10^7 Vm^{-1}, 4 \times 10^7 Vm^{-1})$, azimuthal shift of a particle (green colored sphere) in (i) clock-wise or (ii) anti-clockwise direction occurs at a rotational speed of approximately $3750\ rev\ s^{-1}$without any longitudinal shift. Any particle having radius less than $0.20\mu m$ should rotate successfully and in the plane of rotation the nearest distance between particles is approximately $0.95\mu m$. The maximum intensity in this case is approximately $1.5 \times 10^9\ W\ cm^{-2}$.

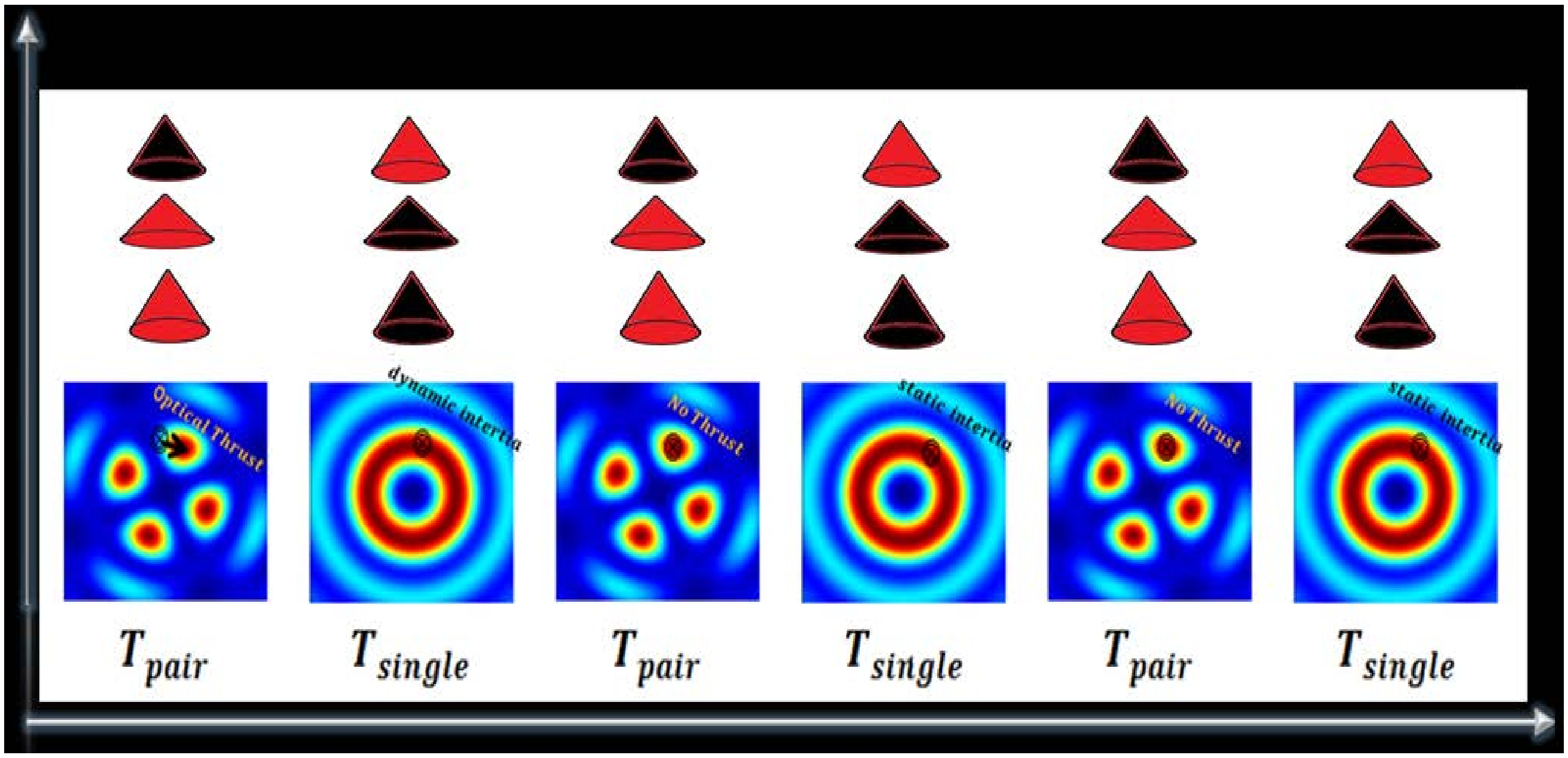

Fig. 6: Trapping a particle at a fixed location using an extra Bessel beam along with existing Bessel beam-pair. In this case, the frequency difference of the two beams is to be made as small as possible. Both the superimposed Bessel beam-pair is to be kept switched-on for the time span of $T_{pair}$ while keeping the third beam switched-off at that time-duration. Then the third beam is to be made switched-on for $T_{single}$ time span while keeping the other beams switched-off. This switching must be done continually $(T_{pair}, T_{single}, T_{pair}, T_{single}, \ldots \ldots \ldots)$and at the starting of each $T_{pair}$ a fixed pattern (shown in the figure) is to be projected and $T_{pair}$ should be small enough so that the dynamic shifting of the starting pattern during $T_{pair}$ is negligible. If the particle is not initially at the desired location, it will experience the optical gradient thrust during $T_{pair}$ and will travel towards the nearest minimum potential point. It will continue travelling during $T_{single}$ along the minimum potential path (elliptical if $\angle i \neq 0^0$) because of dynamic inertia. Once it is placed at the desired minimum potential point, it will not experience any optical gradient thrust during $T_{pair}$. During $T_{single}$ span of time, the particle may slightly dislocate because of very weak gravitational pull but it will be easily relocated during the next $T_{pair}$ time-span with the help of optical gradient force.

$(a)$

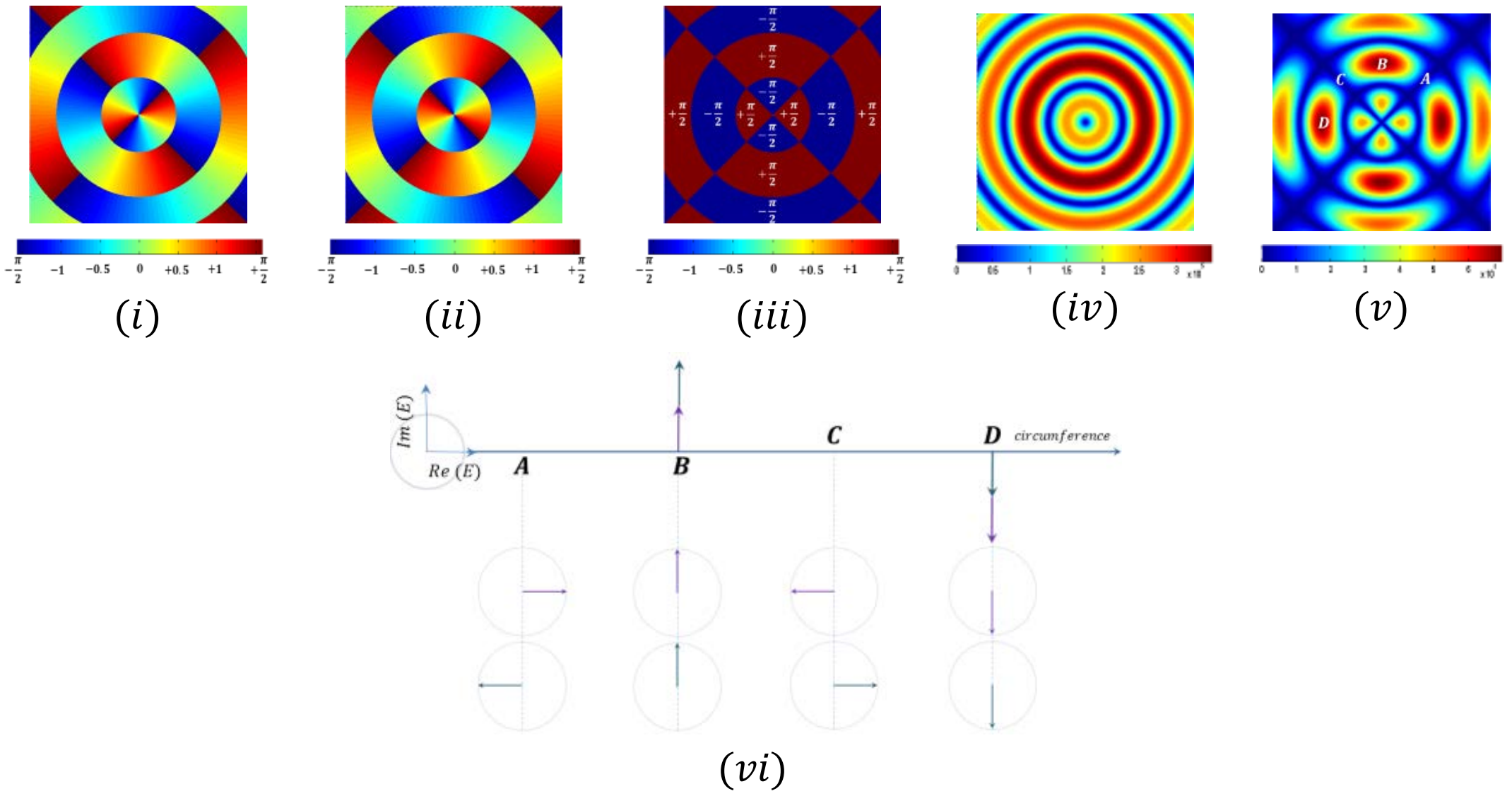

$(b)$

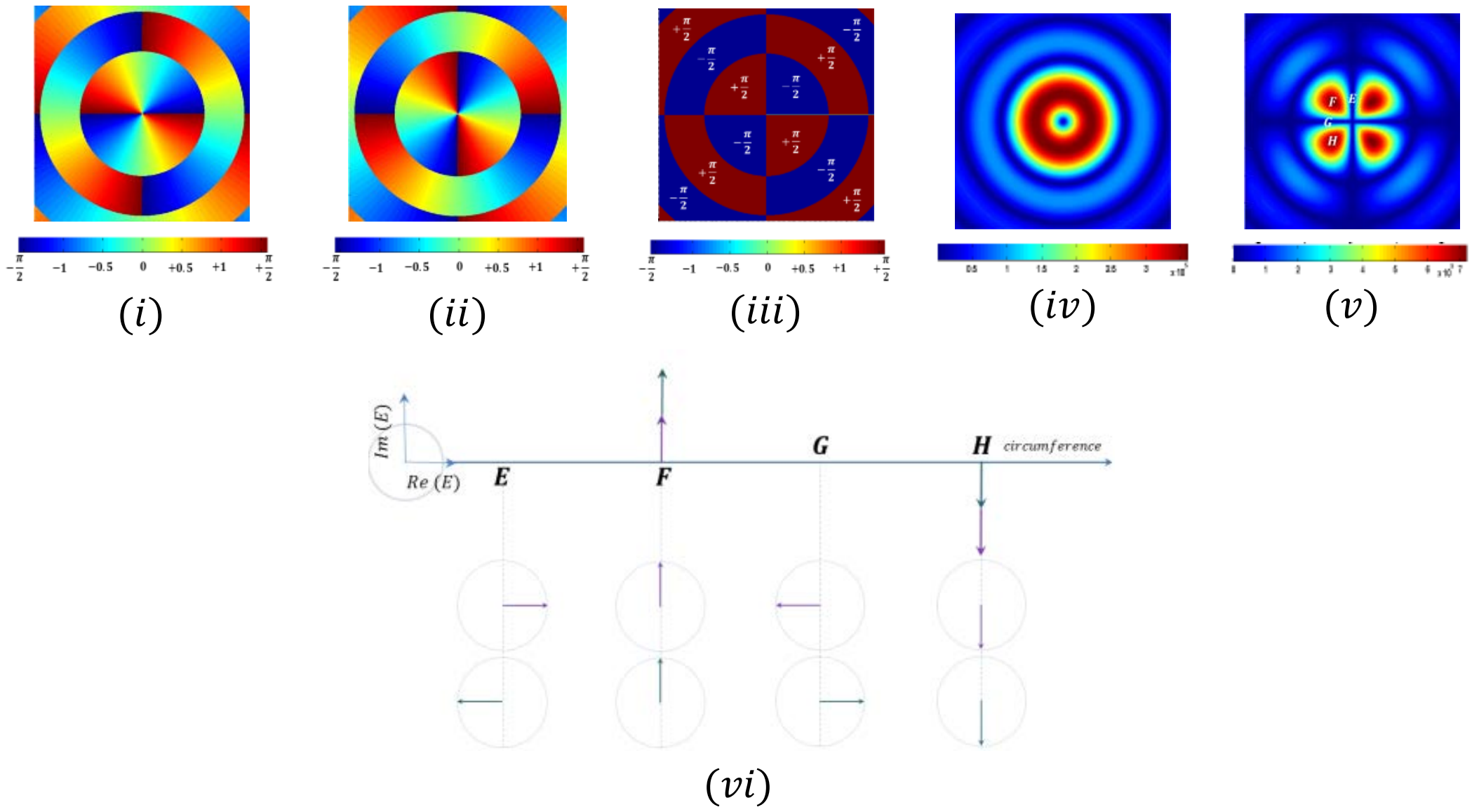

$(c)$

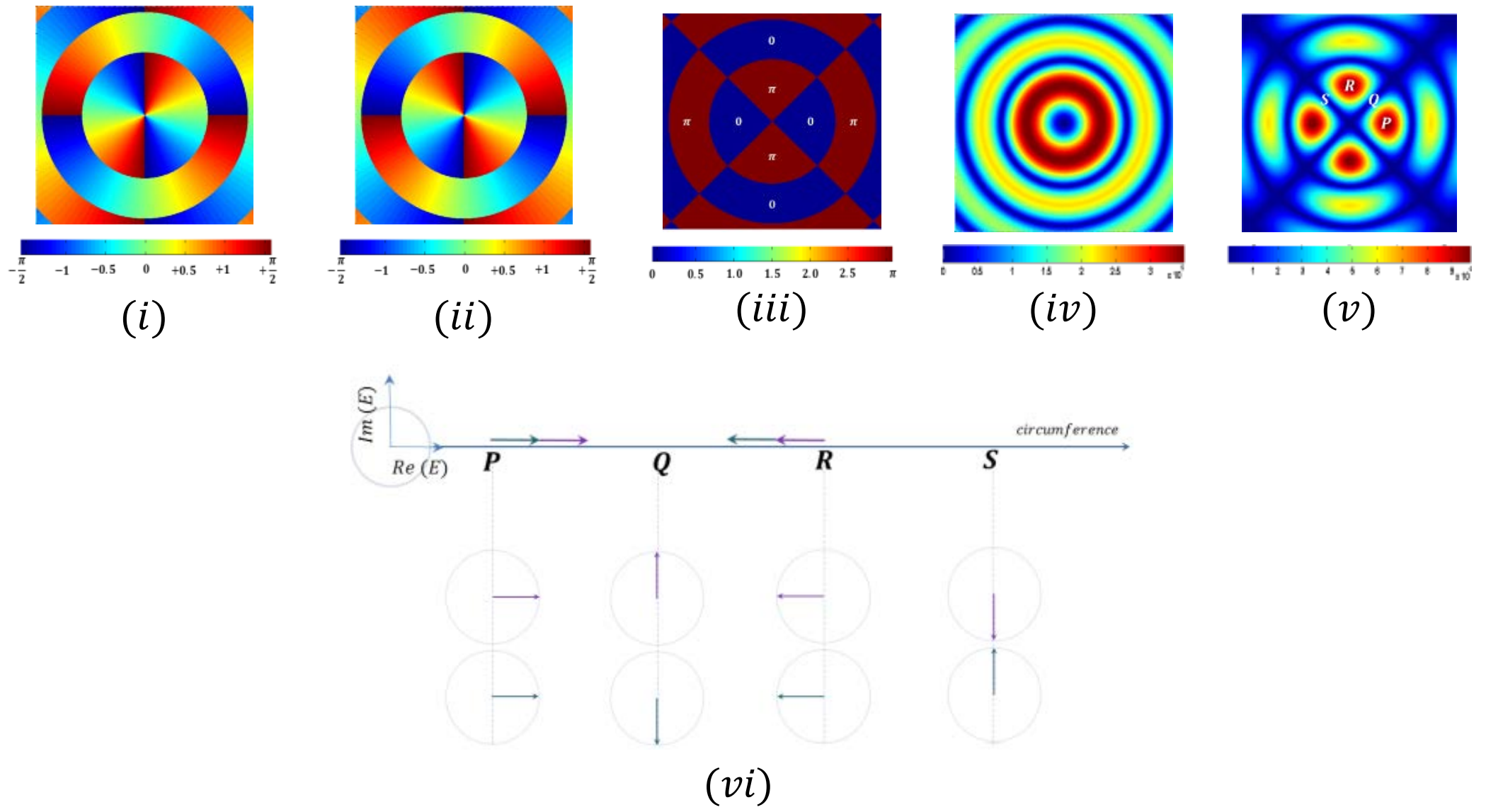

Fig. 7: Individual and resultant phase distribution of electric fields of two Bessel-beams of same but reverse order. The amplitude distribution of the resultant electric field is similar to that of resultant phase distribution which is the reason of creating $|m^{(1)} - m^{(2)}|$ number of maxima or minima in a periodic pattern within $\langle 2\pi \rangle$ range of any cross-section perpendicular to the beam-axis. Here $(m^{(1)}, m^{(2)}) = (+2, -2)$has been taken for simplicity; (i) phase distribution of electric field of first Bessel-beam (ii) phase distribution of electric field of second Bessel-beam (iii) phase distribution of electric field of resultant beam (iv) amplitude distribution of electric field of either of the two Bessel-beams (same in this case) (v) amplitude distribution of electric field of the super-imposed beam (vi) phasor-diagram of electric fields in (a) radial (b) azimuthal and (c) longitudinal directions separately.

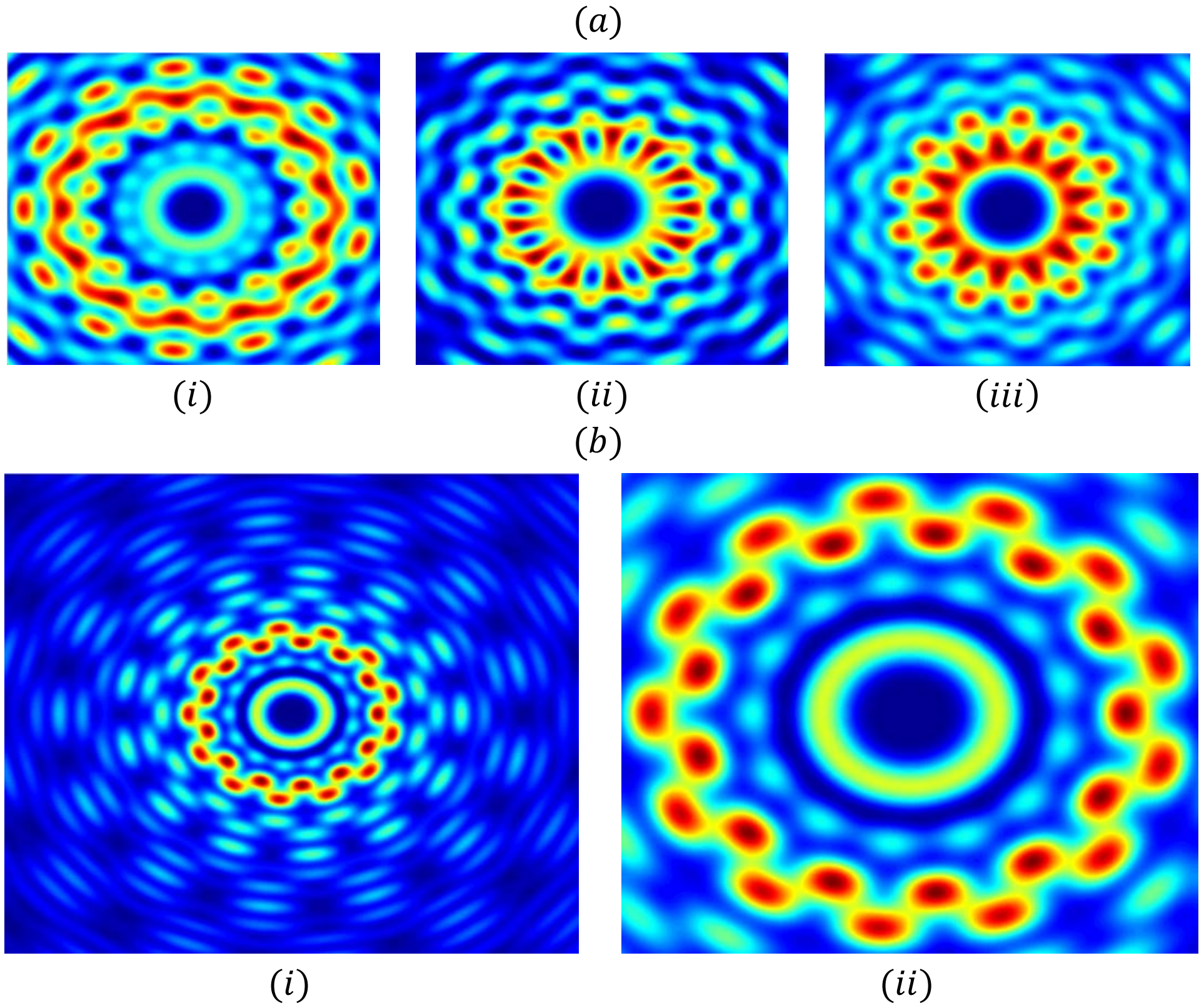

Fig. 8: Optical Manipulation of fully immersed multiple particles and particles of various sizes inside liquid by increasing the intensified regions (binding area of light seeking particles). Superimposed co-axial higher order Bessel-pair having cone-angles of $\theta_{cone}^{(1)} = 45^0$and $\theta_{cone}^{(2)} = 73.57^0$ respectively (at incident angle$\angle i = 45^0$ ) has been used. Deatiled illustrations and exact parametersof multiple particles having different sizes will be found in the main text and in the supplementary. (a) Higher value of $|m^{(1)} - m^{(2)}|$ generates higher number of separate minimum potential regions in a certain cross-section perpendicular to the beam-axis. (i) Bessel orders (+9,-4) in TE mode (ii) Bessel orders (+4,-9) in TE mode (iii) Bessel orders (+4,-9) in TM mode. (b) The intensity profile in transverse plane for beam orders

(+9,-4) in TM mode; capable of binding 26 number of particles in a single transverse plane. (i) Wavelength $\lambda = \sim 1\mu m$ allows optical manipulation of dielectric particles of radius below $150nm$. The refractive index of silica and water at this wave-length is 1.535 and 1.33 respectively. The frame-size is "$17.25\mu m \times 14.4\mu m$" and the minimum inter-distance amongst particles in the inner-binding region is $\sim 1.034\mu m$and that in the outer-binding region is $\sim 1.240\mu m$. (ii) The dimension of intensity profile being directly proportional to the range of wave-length, offers a larger dimension "$69\mu m \times 57.6\mu m$",if the wavelength is increasedto$\lambda = \sim 10\ \mu m$.The maximum radius of the particles, in this case, is $1.5\mu m$ with the inter-distance of $\sim 10.338\mu m$ and $\sim 12.406\mu m$ if located in the inner and outer binding region, respectively. The refractive index of the particle and water at this wave-length is 2.8118 and 1.218 respectively.

$(a)$

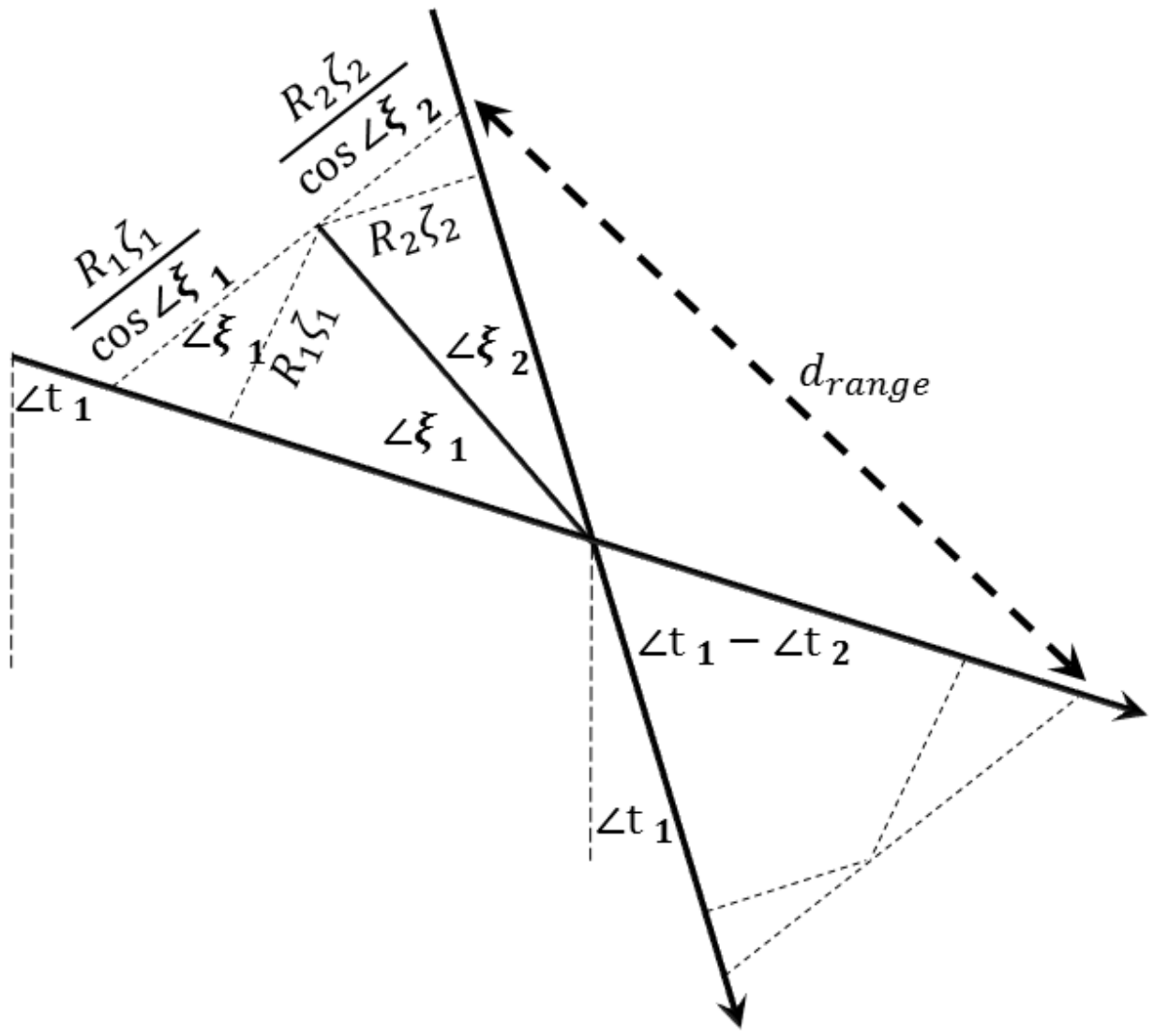

$(b)$

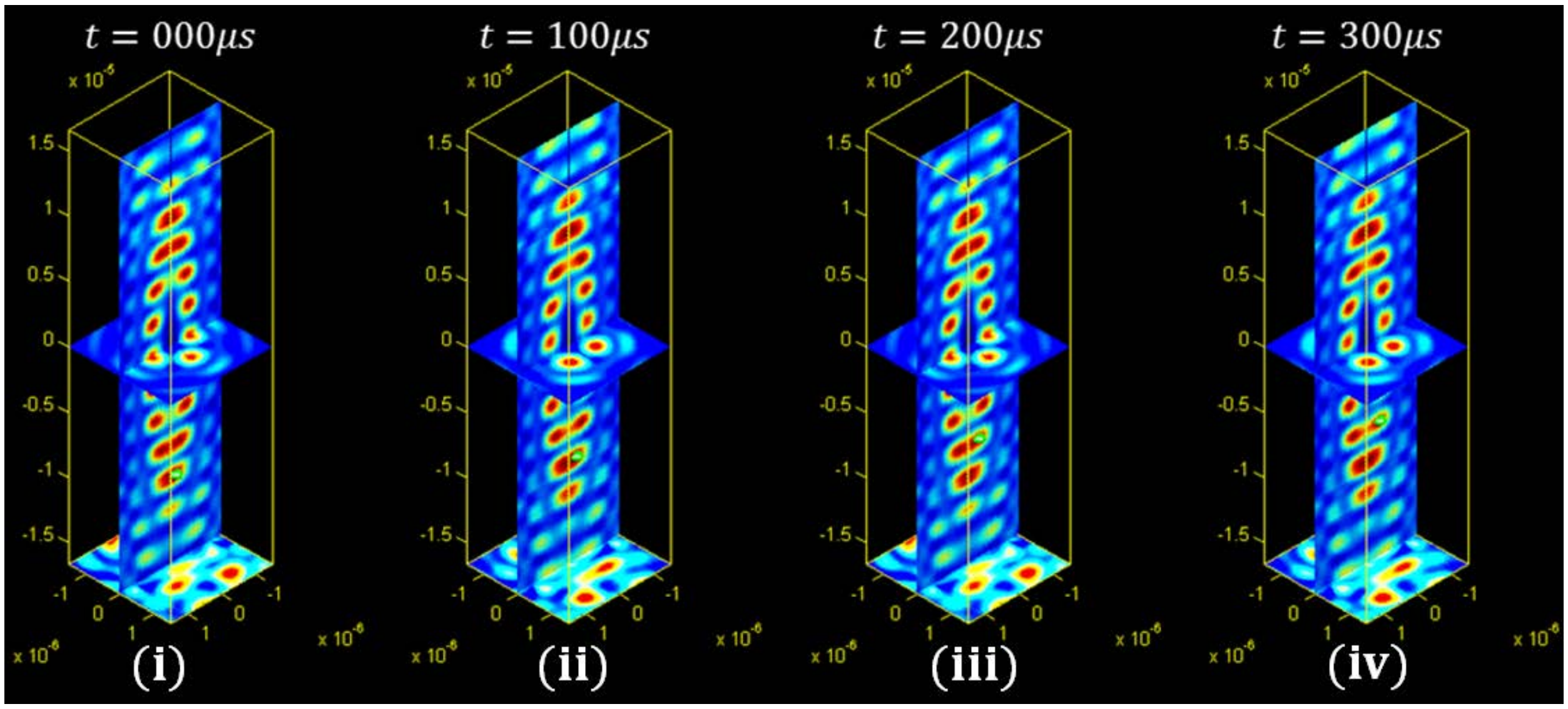

Fig. 9: (a) Geometric figure showing the range and direction of levitation of particle inside liquid medium (also cf. Fig. 1 (c) and (d) and the main text section before conclusion). (b) Levitation of particles located within a certain range only by using non-coaxial Bessel beam-pair having $\left(m^{(1)},\ \theta_{cone}^{(1)},\ \omega^{(1)},\ C_E^{(1)},\ C_H^{(1)},\ \angle i^{(1)}\right) = (+2,\ 50^0,\ 2\pi \times 300\,THz,\ 10^6 V\,m^{-1},\ 0\,A\,m^{-1},\ +5^0)$ and $\left(m^{(2)},\ \theta_{cone}^{(2)},\ \omega^{(2)},\ C_E^{(2)},\ C_H^{(2)},\ \angle i^{(2)}\right) = \left(-2,\ 70^0,\ \omega^{(1)} - 2\pi \times 5\,kHz, 10^6 V\,m^{-1},\ 0\,A\,m^{-1}, -5^0\right)$. The speed of levitation of particle for these parameters is approximately $1.25\,cm\,s^{-1}$ within a range of $20\,\mu m$ approximately. (i) at $t = 0\,ms$ particle is at $z_p = -10.00\,\mu m$ (ii) at $t = 0.1\,ms$ the location of the particle is $z_p = -8.75\,\mu m$ (iii) at $t = 0.2\,ms, z_p = -7.50\,\mu m$ (iv) at $t = 0.3\,ms$ the particle reaches at $z_p = -6.25\,\mu m$. . The density of the particle ($SiO_2$) is $2650\ kgm^{-3}$ and the wavelength being $1\mu m$, the refractive index of the particle is $1.535$.

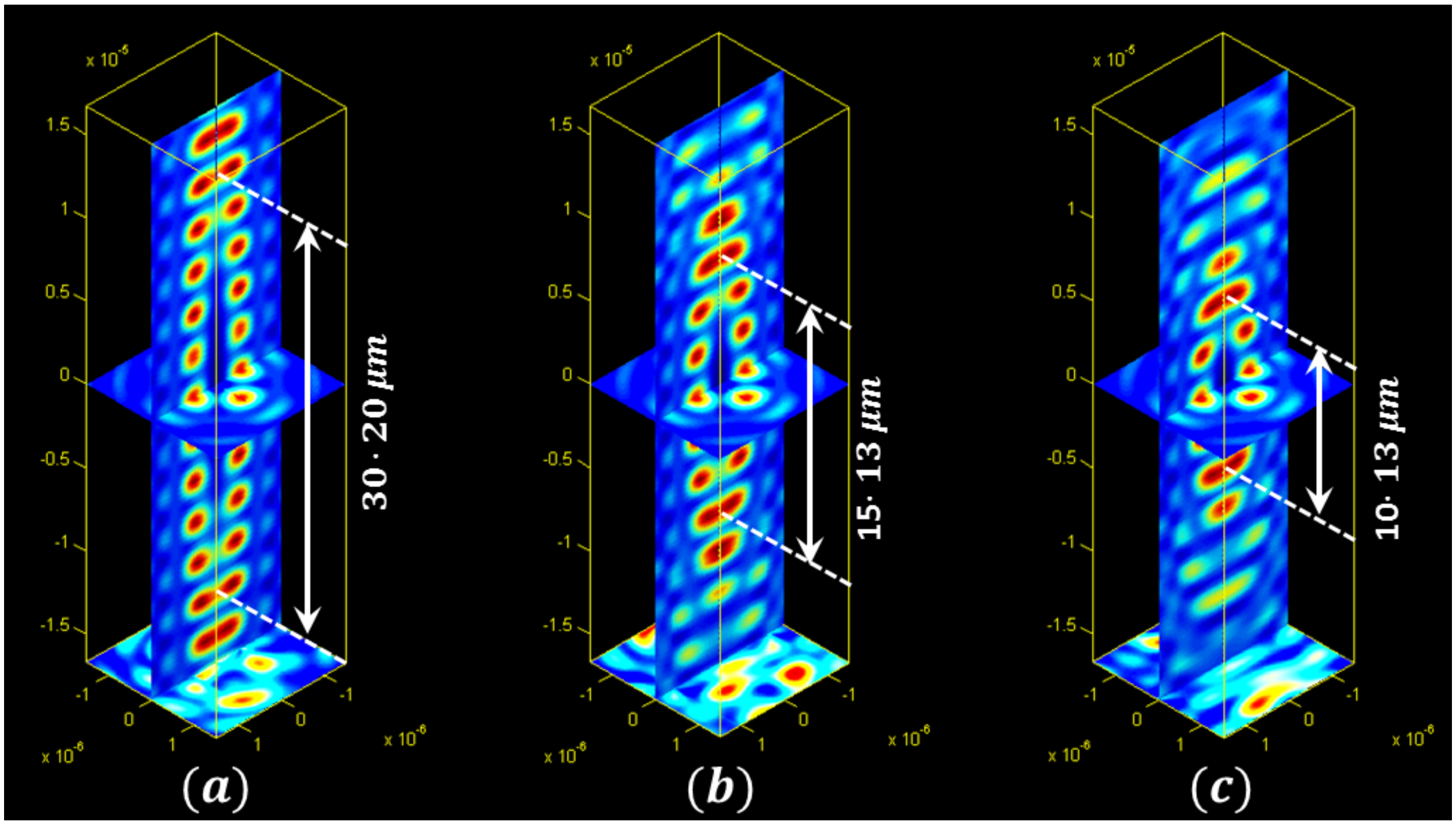

Fig. 10: Depth-wise levitation of particles by changing the incident angles of Bessel beams having $\left(m^{(1)}, \theta_{cone}^{(1)}, \beta_{gas}^{(1)}, C_E^{(1)}, C_H^{(1)}\right) = (+2,\ 50^0,\ 10^6 Vm^{-1},\ 0\ Am^{-1})$and $\left(m^{(2)}, \theta_{cone}^{(2)}, C_E^{(2)}, C_H^{(2)}\right) = (-2,\ 70^0,\ 10^6 Vm^{-1},\ 0\ Am^{-1})$ (a) by projecting beams at incident angles of $\angle i^{(1)} = +2.5^0$ and $\angle i^{(2)} = -2.5^0$on the water surface, particle levitation can be levitated from $15.1\ \mu m$ below the crossing point of beam-axes to $15.1\ \mu m$ above that point (b) at incident angles of $\angle i^{(1)} = +5.0^0$ and $\angle i^{(2)} = -5.0^0$ the levitation-range reduces to approximately $15.13\ \mu m$(c) at incident angles of $\angle i^{(1)} = +7.5^0$ and $\angle i^{(2)} = -7.5^0$ the range of levitation becomes $10.13\ \mu m$.